# Break-junction technique in application to layered superconductors (Review article)

S.A. Kuzmichev[1*], T.E. Kuzmicheva[2]

[1] *M.V. Lomonosov Moscow State University, Faculty of Physics, Low temperature and Superconductivity Department, GSP-1, Leninskie Gory, 119991 Moscow, Russia*

[2] *P.N. Lebedev Physical Institute, Russian Academy of Sciences, Leninsky prospekt 53, 119991 Moscow, Russia*

* e-mail: kuzmichev@mig.phys.msu.ru

Here presented a systematic study of superconductor-constriction-superconductor contacts realized by a break-junction technique in layered superconductors. Depending on the constriction transparency the tunneling and SnS-Andreev spectroscopies could be used, for the direct determination of values of superconducting gaps, characteristic BCS-ratios and gap temperature dependences in cuprate superconductors, magnesium diboride, novel pnictides and chalcogenides. Basing on these data, one can estimate the gap anisotropy magnitude as well as values of electron-boson coupling constants. We discuss the advantages and difficulties of the break-junction technique and demonstrate this method is powerful enough for high-resolution investigation of optical phonon modes in high-temperature superconducting cuprates and for creating of the contacts with the selective transparence in $Mg_{1-x}Al_xB_2$ compounds.

*In memory of Ya. G. Ponomarev, professor, our teacher, warrior, and outstanding experimenter. In memory of S. N. Tchesnokov, mentor, friend, and extraordinary inventor*

*Introduction*

The experimental determination of an extremely important property, the superconducting order parameter, is one of the key problems in superconductivity physics. A simplified but demonstrative description of the physical meaning of this value is offered by the theory presented by Bardeen, Cooper, and Schrieffer (BCS), in which the forbidden energy band (for conventional charge carriers, i.e., normal electrons), which appears as a gap in the dependence of the electron density of states on energy, is considered to be the superconducting order parameter [1]. The accurate and unambiguous measurement of the superconducting gap affects not only the development of the theoretical understanding of the nature and mechanisms of high-temperature superconductivity (HTSC), but also the possibility of developing advanced HTSCs. The spectroscopic methods based on Josephson effects, quasiparticle tunneling, and Andreev reflection effects [2,3] are considered to be the most useful instruments in superconductor studies. Tunneling and Andreev spectroscopy allow us to measure not only the value of the superconducting gap [4–6] but also to study other properties of the material (the electron-phonon, for example see Refs. [7] and [8]), with a high degree of accuracy. This review is dedicated to the unique technique of creating (nano)contacts on a microcrack ("break-junction"). This method has a number of advantages in comparison to traditional tunneling techniques, which allow it to locally determine the bulk value of the order parameter using different spectroscopic methods on the same cryogenic cleaved surface.

This study consists of 5 parts. The first is a brief survey of the main stages involved in creating break junctions, as well as other tunneling techniques. The second provides a detailed description of how the "break-junction" experiment on layered materials is set up and conducted. The third section is a discussion of the Andreev and intrinsic Andreev spectroscopies of the obtained break-junctions; we provide the expected form of the current-voltage characteristics (CVC) and dynamic conductance spectra of the contacts for the case of a multiple gap superconductor as well as the anisotropic order parameter. The fourth section is dedicated to the implementation of the spectroscopic methods considered above in "break-junction" experiments on HTSC cuprates, magnesium diboride and iron superconductors. The fifth chapter briefly presents the main conclusions regarding the benefits of the techniques involved in creating break junctions, and their applicability to a variety of samples.

**1. The history of how the tunnel experiment was developed**

By the time the scanning tunneling microscope (STM) was invented [9,10]., classic SIS and NIS contacts (S is a superconductor, N is a non-ballistic layer of normal metal, I is the insulator) were usually constructed using a natural (oxidation) or artificial (mesa structure) oxide layer between two superconducting electrodes, as well as a vacuum barrier [11–13]. Along with the likes of "solid" tunnel junctions, squeezable contacts [14–18] also enjoyed a lot of popularity, in which the role of the tunnel barrier was played by a gap with a thickness about equal to the interatomic distance between two tightly pressed superconducting electrodes made of the studied material, immersed in liquid helium. Precise adjustment of the contact force between two electrodes changed the contact resistance, which in turn, affected only the amplitude of the tunnel characteristics of the dI(V)/dV-spectrum, without shifting their position and the value of the determined superconducting gap.

In order to implement Andreev reflection spectroscopy,3 Andreev point NS contacts were used, which were created by compressing a metal needle [19,20], or by "pinning" the metal wire to the surface of the superconducting sample using current pulses [20]. This technique is usually referred to as "PCAR" (point contact Andreev reflection). Such tunneling and Andreev contacts have a number of benefits: sufficient mechanical stability for the study of superconducting properties using the corresponding methods of spectroscopy, the opportunity to study samples that are microscopic in size, a controllable configuration (creating a contact along the ab and c-directions of the single crystal lattice), but they also have evident drawbacks: the superconducting order parameter near the surface exposed to degradation is often reduced in comparison to its bulk value (as demonstrated in Refs. [21] and [22], for magnesium diboride, within the scanned surface with an area 0.5·0.5 $\mu m^2$ the amplitude of the quasi-two-dimensional order parameter varied from zero to the maximum value); the point at which the measurement occurs is also the location where the current is injected (since the sample is connected via a three-point circuit), which can lead to an uncontrolled local increase in the temperature. Moreover, the poorly-controlled geometry of the barrier in a NS contact leads to the formation of micro short-circuits and the appearance of hundreds of thousands parallel nanoscale contacts having different configurations (including tunneling) [20,23]. To this day, the question of whether or not PCAR methods are applicable to the study of two-gap superconductors remains unanswered: the proximity effect gives rise to the formation of Cooper pairs with two different binding energies in the metal part of the NS contact, which correspond to the respective values of the superconducting gaps. As a result of scattering by defects, there is a mixing of these induced pairs within the metal, which leads to the convergence of two desired binding energies, up to their unification with a gap magnitude resembling the rms value.

The concept of tunneling break-junctions was first proposed in the early 1980s [24], almost at the same time as the invention of the STM [9,10]. In the simplest configuration the fragile wire from the test superconducting material (in the initial studies it was Nb) is fixed to a flexible substrate, and immersed in liquid helium. At T = 4.2K the mechanical deflection of the substrate split the wire in halves, creating two equal superconducting banks. Further on, by adjusting the deflection of the substrate, the ends of the wire were separated by a few angstroms in order to obtain a SIS tunnel contact. The paper authored by Moreland and Ekin [18] opened a new level in the development of tunnel research: this group was able to create experimental techniques that not only preserved all advantages of the classical methods, but also practically eliminated the shortcomings thereof. According to estimates in Refs. [18] and [23], the mechanical stability of "broken" contacts did not yield to the strength of the squeezable contacts, structures with natural oxide barriers or STM contacts. At the same time the experimenters solved two important problems that are typical for classical SIS, NIS, and NS contacts:

(1) significant weakening of studied surface degradation due to the creation of a cryogenic cleavage in an inert atmosphere, which ensured a high purity of superconducting electrodes;

(2) an absence of mechanical pressure on the contact area, which prevented the distortion of the superconducting properties of the material.

Several years after Moreland's study was published, more sophisticated "break-junction" setups were created, which implemented both rough and fine adjustment of the substrate deflection, as well as systems that allowed the researchers to work in vacuum with non-brittle materials (such as simple metals) and films [23,25–27]. The main feature of this new generation of break junctions was the ability to transfer the contact between the low capacity (so called "constrictions" [23]), high capacity ("tunneling" contact [23]), and intermediate regimes using mechanical regulation of the tunnel barrier thickness. Later the technique of creating break junctions for the implementation of SIS and SnS spectroscopy was successfully applied in studies pertaining to the properties of HTSC cuprates [8,28–45] (for a review, see Refs. [46,47]), magnesium diborides [36,48–59] (for a review see Refs. [60,61]), iron pnictides and chalcogenides [53,62–77].

At the present time, regardless of the obvious advantages over the traditional tunneling methods, the "break-junction" technique is used quite rarely. This is most likely based on a lack of application possibilities, and the prospect of commercializing this technique for superconducting materials usage. In particular, experiments with SnS Andreev break-junctions (high transparency of barrier) are carried out only by our group. However, in 1997 Reed et al. suggested the idea of implemented break-junctions (based on mechanically controlled contact configuration used by Muller et al. [23]) in which broken electrodes were weakly bound by individual molecules [78]. Thus, the unique opportunity provided by "break-junction" technique to create ultrapure break junctions has found wide application in molecular electronics: in particular, in studies of vibrational and transport properties of separate organic molecules (see Ref. [79] for a review).

## 2. The break-junction technique

### 2.1. The configuration of a "break-junction" experiment

Our group uses the "break-junction" technique that was perfected by Ponomarev [28,29], with respect to layered superconductors. In these materials the "break-junction" technique allows for the realization of four methods of investigating the superconducting order parameter on cryogenic cleavages of the same sample: tunnel (Josephson), intrinsic tunnel [80], SnS Andreev, and intrinsic Andreev spectroscopies [81].

A layered single crystal (or polycrystalline sample with crystallites oriented along the c-direction) superconductor is prepared using a thin rectangular plate parallel to the ab-plane of the crystal lattice, with sizes of about $(3–5) \cdot (1.5–2.5) \cdot (0.1–0.4)$ mm$^3$. For especially solid samples ($MgB_2$, for example), it is necessary to also make a notch which demarks the sample into two squares and serves as an additional stress concentrator. An insulating substrate (3) with four copper contact pads (5) is fixed on the U-shaped spring measuring table (8 on Fig. 1) made of beryllium bronze 0.2 mm thick. The substrate also has a sufficiently deep cross-section, which is the mechanical stress concentrator. Along the edges the substrate is further fixed to the table using a bandage (4). Two current and two potential contacts are fed to the pads in order to implement a standard four-point measurement scheme. Sample (7) is fixed in the middle of the contact pads using massive drops of indium-gallium solder (6) (spreadable paste) that is liquid at room temperature, at the sample corners. The use of a eutectic protects the thin plate of the sample from premature breakage at inevitable deformations of the substrate during the process of mounting and at the initial cooling of the insert. The configuration of the measuring table can also be successfully used for mounting the whisker single crystals up to 5mm long, as shown in Fig. 1(c): in order for a whisker not to "fall down" into the gap between the copper pads, a substrate made out of tissue paper is used, and four In-Ga contacts are applied parallel to each other and perpendicular to the single crystal needle across the length of the sample.

After fixing the table to the insert, a screw with a micrometer thread (1) is applied to the former; indentations are made at the end of the micro screw and at the center of the table, into which the needle is inserted tightly (2). The needle is used to transfer only the translational displacement of the microscrew to the surface of the table. For the successful creation of cryoge-nic cleavages it is extremely important to ensure that the sample is not prematurely cracked during cooling. Therefore, it is necessary to slightly strain the table spring using the needle in advance (by making a half-turn with the microscrew, for example), and gradually weaken the deformation during the cooling process. Microcracks are created in the crystals at helium temperatures by applying precise mechanical pressure to the measuring table, which bends the substrate made of polyepoxides bonded paper (FR-2 or getinax) with the sample along the stress concentrator (see Fig. 1(a)); at small displacements this corresponds to the sample split in a direction perpendicular to the concentrator. The layered sample is held firmly by frozen In-Ga solder, and stratified in the stress concentrator region along the ab-plane. In general, two cryogenic cleavages are created in the sample, which represent steps and terraces separated by a weak coupling region, i.e., we get an ScS contact (where c is a constriction). In the experiment the c-region can formally exhibit properties of insulator (I), a normal metal (N), or of a thin (in comparison to the length of the carrier mean free path) normal metal (n), depending on the barrier transparency Z. This parameter can be regulated in STM experiments [82]. Given a minimum substrate deformation the current across the break-junction flows along the crystallographic axis c.

It is well-known that the surface of many layered HTSCs either does not carry any information about the bulk properties of the material (for example, HTSC cuprates are sensitive to oxygen doping loss, iron pnictides LiFeAs are highly susceptible to degradation in the presence of water vapor, etc.), or hinders the process of conventional surface techniques, for example, due to the formation of a Schottky type barrier (so in oxypnictides LnOFeAs, Ln is a lanthanide and the surface turns out to be charged [83]). The values of the gap and the critical temperature $T_c$ on the surface of the crystal can be markedly different from those in the bulk of the sample. In our experiments on layered materials, in contrast to the procedures proposed in Refs. [23] and [24], in the process of creating microfractures the superconducting banks of the sample are not separated by a significant distance; the microcrack is formed in the bulk of the sample and is not visible on the surface. Given this type of the "break-junction" technique, the cryogenic surfaces are naturally protected from degradation caused by the penetration of "dirt" from the atmosphere into the sample, precisely because they are in the bulk; the cleavages remain as clean as possible. We identify the onset of the crack dividing the sample plate into two halves in real time according to the appearance of a slope along the current-voltage characteristics, at currents of about 5–10 mA. We observed that given a considerable separation of the sample halves and a full opening of the cryogenic cleavages, the surface degrades, which leads to an increase of both normal scattering Γ, and a decrease in the amplitude of the superconducting gaps Δ. The emergence of Schottky-type barriers is sufficiently rare and is easily controlled according to the preservation of the CVC symmetry.

Break-junctions are located between massive superconducting banks of the sample, which ensure reliable heat sink from both sides (in contrast to NS contacts and especially mesa structures). Moreover, since the microcrack is located away from the potential and current contacts of the sample, then the heat generated by the latter does not lead to the heating of the studied point. Therefore, the experiment configuration we used almost completely eliminates the chemical, thermal, and mechanical affecting on the region of the ScS contact. The resulting values of the superconducting gaps, therefore, are as close as possible to the bulk value of the order parameter. According to our estimates, the radius of the break junctions a ~ 1.2–30 nm [51,71], thus, the superconducting properties are studied locally (i.e., within the contact area). In particular, this makes it possible to accurately determine the local critical temperature $T_c^{local}$, which is the transition temperature of the contact area into the normal state. In the experiment, $T_c^{local}$ is determined according to when the dI(V)/dV-spectrum becomes linearized (which corresponds to the ohmic CVC). Due to the natural inhomogeneity of the samples, the local critical temperature could be significantly different from the average over the bulk of the sample $T_c^{bulk}$ (determined, for example, in resistive measurements or the temperature dependence of the magnetic susceptibility of the sample). Knowing $T_c^{local}$ can help us determine the real (local) value of the characteristic $2\Delta/k_B T_c$ ratio, which is the key to the fundamental conclusions as to the nature of superconductivity in a given material.

It is known that steps and terraces form on the cryogenic cleavages of any layered compounds. The height of these steps is a multiple of the lattice parameter c; for cuprates it is c/2 due to the half-period "shift" of the lattice structure in the ab-plane. Given a small deformation of the substrate, the cryogenic cleavages slide relative to each other with precision along the terraces (in the ab-plane), which implies the creation of a tunneling current through the fracture in the c-direction, and enables a smooth and reversible mechanical regulation of the contact area. On the contrary, if we assume that the superconducting banks touch along the c-direction (i.e., there is a tunneling current in the ab-plane), this contact would not be subject to a smooth regulation: a backtrack of the banks would cause an abrupt increase in contact resistance. As such, the orientation of the contact (j||c) can be accurately determined throughout the experiment. The steps and terraces of the cryogenic cleavages often exhibit properties of stack contacts such as ScSc-…-S. Since they in fact are natural mesa structures, such arrays are electrically equivalent to a chain of series-connected identical (single) ScS contacts. Thus, the position of the singularities caused by the bulk effects on the CVC and the dynamic conductance spectrum of stacks made of m contacts (m is a natural number) will be multiplied by an m times in comparison to the I(V) and corresponding dI(V)/dV of a single contact. The "break-junction" technique can be used to achieve both tunnel SISI-…-S, and Andreev SnSn-…-S structures.

A unique advantage of the break-junction is the possibility of its precise regulation during low-temperature experiments, i.e., studying the properties from one point on the surface of a cryogenic cleavage, to another. Mechanical readjustment can be used to obtain dozens of single and stack contacts in the same sample by scanning cryogenic cleavages, which brings the possibilities of this technique closer to those of STM, in this respect. A set of large statistical data allows us to confirm the absence of any dimensional effects influence on the results of the studies (since the dimensions and normal resistance of each contact are random) and to assess the uniformity of the superconducting properties of the sample.

*2.2. The methods of recording the dynamic conductance spectra of tunnel junctions*

Using a source of current rather than voltage when studying tunnel junctions implemented on superconductors solves at least two problems: firstly, it gives the opportunity to record supercurrent at zero bias, and secondly, if we have the appearance of (ohmic) contacts connected in parallel to the junction under the study (which is typical for "break-junction" technique), the dynamic conductance spectrum dI(V)/dV shifts strictly vertically (i.e., the bias of any spectrum singularities remain unchanged), which is very important for the accurate determination of the superconductor energy values. For the same reasons a hardware-based determination of the dI(V)/dV-dependence instead of dV(I)/dI is also preferable.

In order to measure the dI(V)/dV characteristics we use the standard current modulation technique and a hardware control system over the measuring bridge balance. The source of the current fixes the current through the sample, while the DC gets mixed with a small amplitude AC having a frequency of about 1 kHz from the external oscillator. The multiplying digital-to-analogue converter (DAC) mounted on a digital IO board can scale the current modulation amplitude, while the automatic computer tracking system using the bridge unbalance signal coming from a selective nanovoltmeter (lock-in amplifier) controls the multiplying DAC and withholds the response modulation amplitude for voltage equal to a certain reference amplitude from the same generator of the sinusoidal signal. The scaling factor of the modulation signal amplitude (with respect to current) is established using the final balance of the electric bridge, recorded in the multiplying DAC, and is proportional to the slope of the CVC at the measured point. It is only after the system has determined the CVC derivative at the desired point that the current through the sample is changed, and the measurement cycle is repeated. By measuring the second harmonic of the modulation signal, it is also possible to obtain the second derivative $d^2I(V)/dV^2$ using hardware.

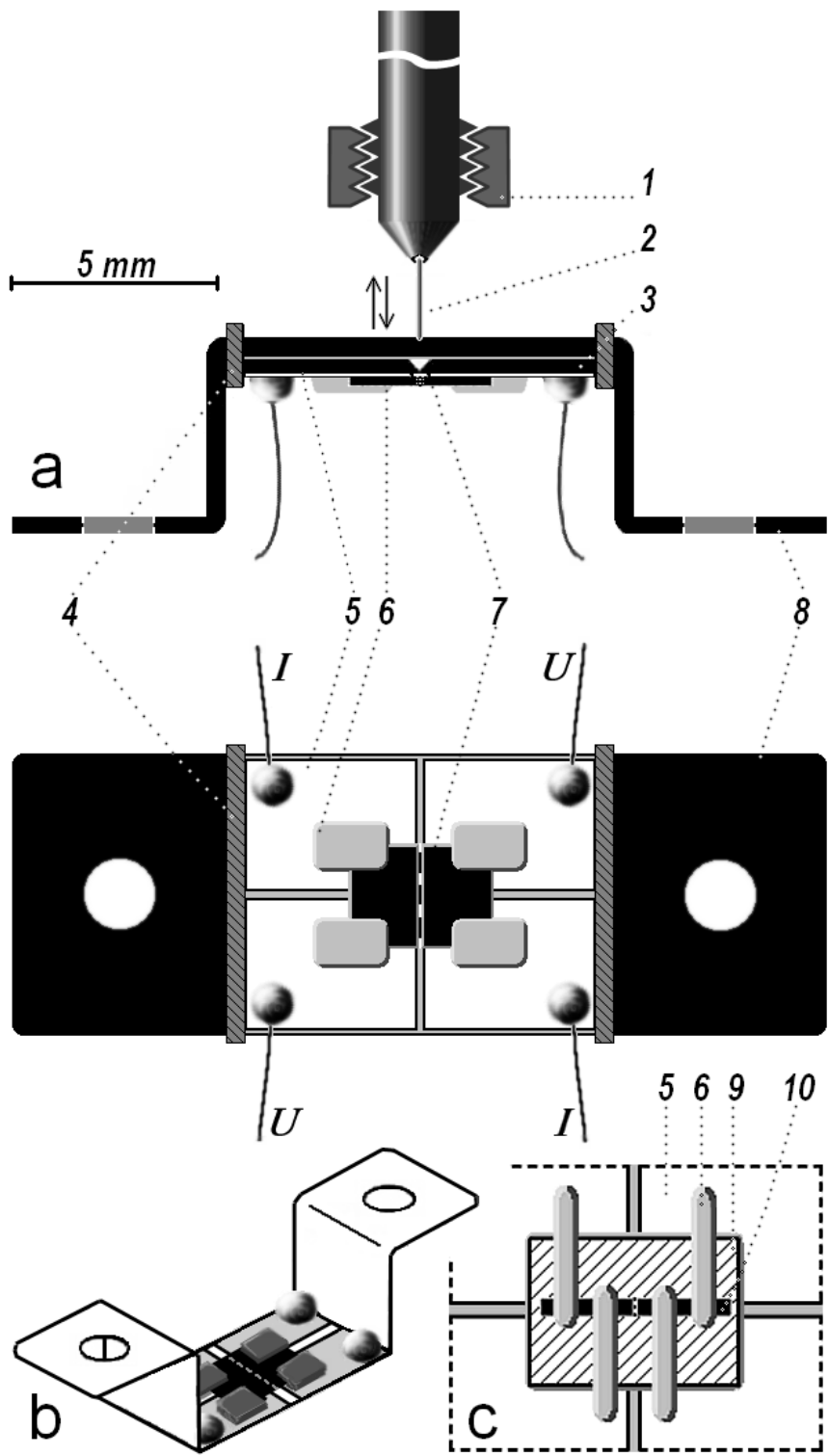

**Fig. 1.** *The design of the table and a diagram of the sample mounting onto the substrate using a four-point scheme for obtaining break-junctions (a). The general form of the table with the sample attached (b). A mounting example of a whisker onto the table contact pads (c). 1 is the screw with a micrometric thread, 2 is the needle for transmitting translational displacement, 3 is the insulating substrate, 4 is the bandage, 5 are the copper contact pads, 6 is the indium-gallium solder, 7 is the sample, 8 is the spring table made out of beryllium bronze, 9 is the flexible insulating lining, 10 is the whisker.*

*2.3. On the possibility of creating break junctions in polycrystalline samples of layered compounds*

The process of forming microcracks in the non-oriented polycrystalline compound that has a layered structure and a significant in-plane to out-of-plane strength anisotropy, is shown schematically in Fig. 2(a). As a result, cracks within the polycrystalline sample can form grain-grain contacts (mainly for crystallites with ab-planes perpendicular to the direction of the fracture; crystallite No. 1 in Fig. 2(a)) and as junctions that are formed by terrace contact of split crystallites, the ab-planes of which are about parallel to the direction of the cleavage (crystallite No. 3 in Fig. 2(a)). In general the likelihood of crystallite fracture depends on the ratio of the mechanical coupling strength between the crystallites $P^{ig}$ and the coupling between the ab-layers of the lattice $P^{il}$. Strictly speaking, the probability that a crystallite (grain) will be split depends on both the $P^{ig}/P^{il}$ ratio and of the spatial orientation of the ab-plane of that grain relative to the plane of the microcrack formation. If the value $P^{ig}/P^{il}$ increases slightly, the share of the split crystallites increases significantly. For example, if $P^{ig}/P^{il}$ increases from 1.1 to 1.2, nearly 2 times as many randomly oriented crystallites will split (the corresponding area is highlighted by hatching in Fig. 2(b)).

It is obvious that for polycrystalline samples synthesized by re-annealing, we should expect a high percentage of stratified crystallites. A simple calculation shows that already at $P^{ig}/P^{il} = 1.1$ we should expect the cleavage of up to 6% of the stratified grains (solid line on Fig. 2(b)), whereas at $P^{ig}/P^{il} = 2.5$ about half of the grains will split. The image obtained at the cleavage of the polycrystalline $Sm_{0.7}Th_{0.3}OFeAs$ using an electron microscope (Fig. 2(c)), clearly shows the steps and terraces at the surface of the split crystallite. Our estimated diameter of the ScS break-junction $2a = 10–60$ nm is several orders of magnitude less than the average grain size and the average width of the terraces (~100–200 nm), such a contact will not yield to a contact created in a single crystal. Moreover, the use of "break-junction" technique in polycrystalline samples is preferable since the crystallites tend to be more homogeneous than single crystals with millimeter dimensions, especially those synthesized under conditions of temperature gradient and/or pressure.

Is it possible for stack ScSc-…-S structures to form in polycrystalline samples of layered superconductors? The attempt to explain the serial ScSc-…-S contacts that are obtained in the experiment as chains of crystallites connected through grain boundaries, and not using intrinsic effects (which are implemented in natural stack structures), cannot withstand criticism. Due to the lack of equivalence between such borders the resistance of these stacks will not scale with number of grains m, thus, the position of the main gap singularities will be random, instead of being a multiple of 2Δ/e; the shape and fine structure of the singularities will not reproduced given a mechanical readjustment of the junction. Moreover, as the number of grains m and intergranular boundaries having non-equivalent resistance in the normal state increases in the chain, the severity of the singularities in the spectra will drop dramatically. In our experiments we observed quite the opposite: in both single crystals and polycrystalline samples of similar compounds a typical resistance of ScS contact was reproduced, and the singularities of the dynamic conductance became sharper with increasing m [71]. Note that the position and shape of the singularities along the dI(V)/dV spectra (caused by bulk effects such as gap and phonon singularities) are reproduced when scaling the bias of the electric potential by a natural number m in order to normalize the conductance features to a single junction spectrum [36,51,66,68,69,71,74,75,77], and coincide with the single contact characteristics. Similar data were obtained for single crystals of layered superconductors [30,67,72,76,81].

Thus, we can assume that the quality of the spectra obtained on stack contacts increases due to a decrease in the contribution from surface defects to the dynamic conductance of a break junction. It should also be noted that an important feature of these tunnel structures in comparison to artificially created mesa structures, is the minor and controllable influence of local overheating effects due to the significant remoteness of the current injection site into the stack, and the good heat sink on both sides of the array.

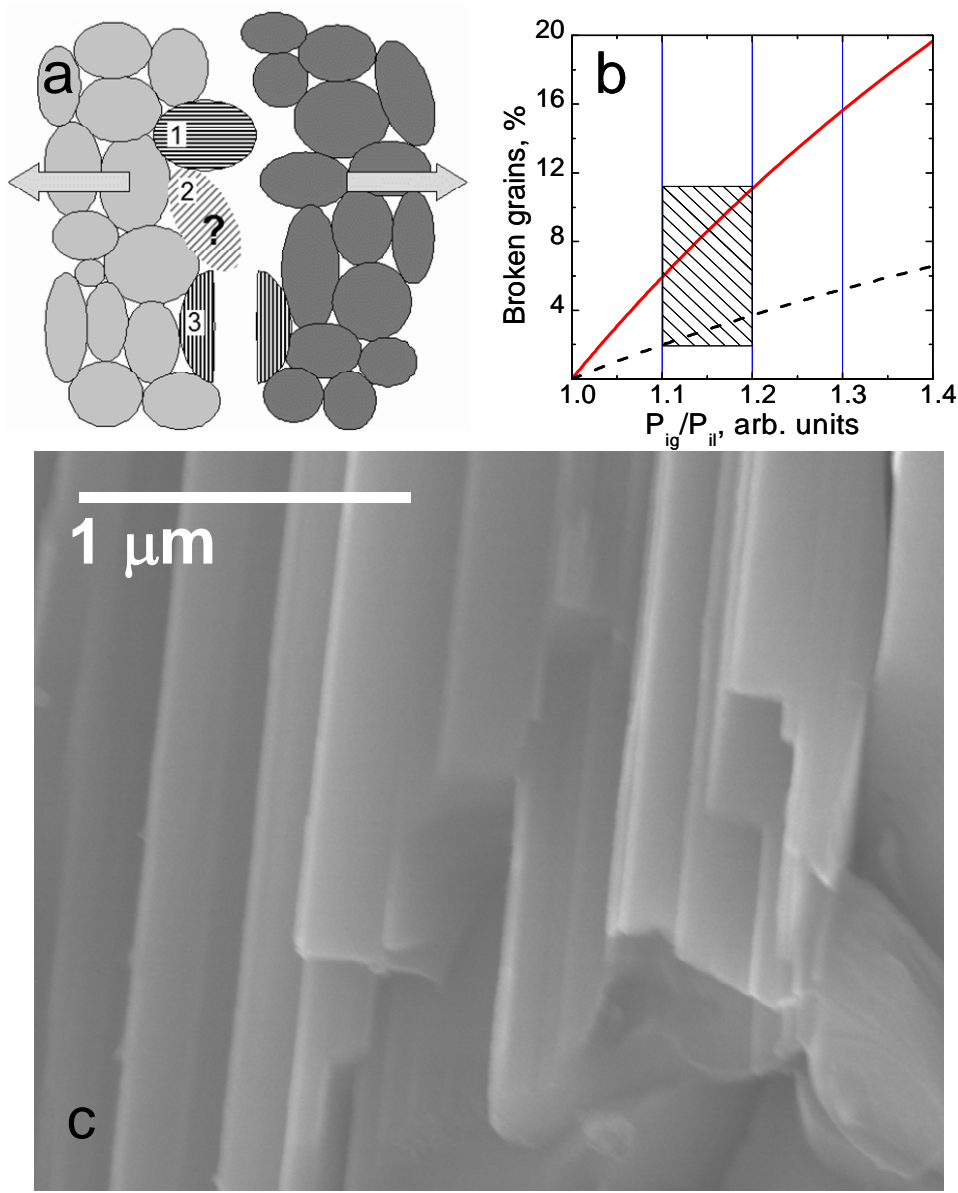

***Fig. 2.*** *A diagram of microcrack formation in the polycrystalline sample. The hatching shows the direction of the ab-planes in the crystal grains (a). The dependence of the part of split crystals on the ratio of the intergranular and interlayer bond strength $P^{ig}/P^{il}$ for samples in which the crystallites have a mechanical bond with all neighboring grains (solid line) and with half the neighboring grains (for loose samples, for example; dashed line) (b). The steps and terraces on the surface of the ruptured crystallite on the cleavage of the layered polycrystalline $Sm_{1-x}Th_xOFeAs$; the image is obtained with an electron microscope (c).*

## 3. Andreev and intrinsic Andreev spectroscopy of superconductors

*3.1 The effect of multiple Andreev reflections*

It is well known that the effect of multiple Andreev reflections [3] of electron is observed in SnS contacts (wherein n is a thin normal metal), when they are ballistic, i.e., when their diameter 2a is less than the mean free path of the carriers *l*. Ballistic contacts are also called Sharvin contacts, named after Yuri Vasilievich Sharvin, who described the physical properties of point contacts with a diameter 2a < *l* in Ref. [84]. The resistance of such a contact is approximately *l*/a times greater than the ohmic and is equal to R = 4ρ*l*/(3pa$^2$) [84]. Note that the lower limit of the 2a range, in which the physics of such point contacts is fully implemented, is a Fermi electron wavelength $\lambda_F$ < 2 nm. Therefore, at 2a < $\lambda_F$ the electron-wave interactions start to play a deciding role; these types of contacts are usually called quantum ballistic. In contrast, in the SnS contacts considered by us, $\lambda_F, \xi$ < 2a < *l* is fulfilled (wherein ξ is the superconductor coherence length) and the ballistic case of electron flight through the contact is realized, which consists of two SN interfaces. For each Andreev reflection from an SN interface the quasiparticle provides a transfer of Cooper pair through the metallic layer, which leads to the appearance of an excess current on the CVCs of Sharvin contacts.

Fig. 3 shows the theoretical dynamic conductance spectra (in reduced coordinates) and the CVC of a SnS Andreev contact, calculated based on the different models for a single-gap superconductor with an isotropic gap (s-wave symmetry). The contact resistance is assumed to be unity. The dI(V)/dV-spectrum for SnS contacts was first calculated by Octavio with Tinkham, Blonder and Klapwijk as coauthors in Ref. [6] (the so-called OTBK model); the corresponding spectrum (T= 0, barrier transparency Z = 1) is shown in Fig. 3 with a black dashed line. Octavio et al. Qualitatively demonstrated that on the dI(V)/dV-spectrum, the effect of multiple Andreev reflections causes a subharmonic gap structure (SGS), which is a series of dynamic conductance minima at bias:

$V_n = 2\Delta/en$, wherein n is a natural number.         (1)

It is obvious that the position of the Andreev features $V_n$ is linearly dependent on the inverse subharmonic number 1/n.

In more recent theoretic studies [85–87] it was shown that at low bias on the CVC of the SnS contact a significant excess of quasiparticle current was observed, the conductance was several times higher than normal, and the I(V)-dependence of such a contact tends toward ohmic at V >> 2Δ/e (see Fig. 3). The CVC region at V→0 is referred to as "foot." We note for the sake of comparison that when one is dealing with NS contact conductance (which is typical for point contact techniques) at low biases there is an expected two-fold increase of normal ohmic conduction, which occurs at bias above the gap values V>Δ/e [5].

The SGS along the dI(V)/dV-dependence for an SnS contact with high transparency (95%–98%), a diameter 2a, and subject to the condition $\lambda_F < 2a < l$, is a series of conductance minima [85–87]. The spectrum calculated according to the Arnold model [85] is shown on Fig. 3 in a thick, light-green line (T=0, probability of overcoming the barrier $T^2$=0.83). The Averin-Bardas model [86] and the calculations performed by Cuevas et al. that are based on it [87] predict a weakly expressed first gap singularity with n=1, and subsequent subharmonics n=2,3,… are sufficiently intense minima, with positions that are also described by Eq. (1). The results of studies conducted by Arnold [85], and Averin-Bardas [86] for ballistic SnS contacts with a high transparency are in great agreement between one another both in terms of predicting the type of singularities of the subharmonic gap structure along dI(V)/dV, and in terms of general exponential course of this relationship at V→0.

The model composed by Kümmel et al. [88] also considers the ratio of the mean free path of the carriers to the contact diameter $l/2a$, and the existence of Andreev quasiparticle band within the gap. A curve corresponding to the case of $l/2a$=5 and T=0.8$T_c$ is obtained using numerical differentiation of the CVC from Ref. [88] and is shown by a dash-dot line in Fig. 3. The presence of such an Andreev band leads to the appearance of satellite minima that accompany each Andreev subharmonic. We will not go into the details of this result from Ref. [88], but will only note that the intensity of the Andreev minima decreases as the number of the singularity n increases, and the amount of observed subharmonics corresponds approximately to the ratio n ≈ $l/2a$. One of the most important conclusions from Ref. [88] is that the position of the SGS minima continues to follow the above formula for Vn at any temperature $0 < T < T_c$. As such, the multiple Andreev reflection spectroscopy method allows us to determine the value of the superconducting gap directly from the positions of the Andreev subharmonics without additional calculations and fitting of the dI(V)/dV-spectrum. This significantly increases the accuracy of determining the superconducting order parameters and the accuracy of the experimental data for two-gap superconductors in comparison to SIS or N(I)S contact spectroscopy, which involve fitting the dynamic conductance by using adjustable parameters [4,5]. Consequently, the temperature dependences of the superconducting gaps are preferably obtained using SnS contacts, which gives us the possibility of accurate estimating the electron-phonon (in general, the electron-boson) interaction constant [52,53].

We would like to underscore that the CVC and dynamic conductance spectra of SnS contacts with a high transparency and foot at V→0, the preparation of which is typical for the "break-junction" method, are different from I(V) and dI(V)/dV of the quantum point contacts with a low transparency and a current deficit at V→0. In Refs. [86,87] it is shown that as the transparency decreases the series of minima turns into a series of maxima. If the diameter of a contact with a high transparency becomes comparable to the length of the mean free path ($2a \approx l$) then SGS become blurred: one observes a limited number of features all the way up to a single one at $V=2\Delta/e$. In the case of superconductors such as cuprate HTSCs, with hole-like Cooper pairs, the mechanism of multiple Andreev reflections does not change.

An important feature of SnS contacts in the c-direction on layered superconducting materials is the fact that since the Fermi surfaces for both electron and hole bands are usually slightly corrugated cylinders, the normal current carriers pass toward the SN interface almost tangentially. This is not typical for classical theories describing the CVC of Andreev contacts, and in the case of Z<5 the probability of the quasiparticles having a normal reflection from the SN interface should noticeably increase, which is defined as $Z^2/(Z^2 + \cos^2\alpha)$, wherein $\alpha$ is the angle to the normal. Note that at $\alpha \to \pi/2$ the value of $\cos^2\alpha$ is proportional to the bias voltage of the contact. Of course this case requires further theoretical study.

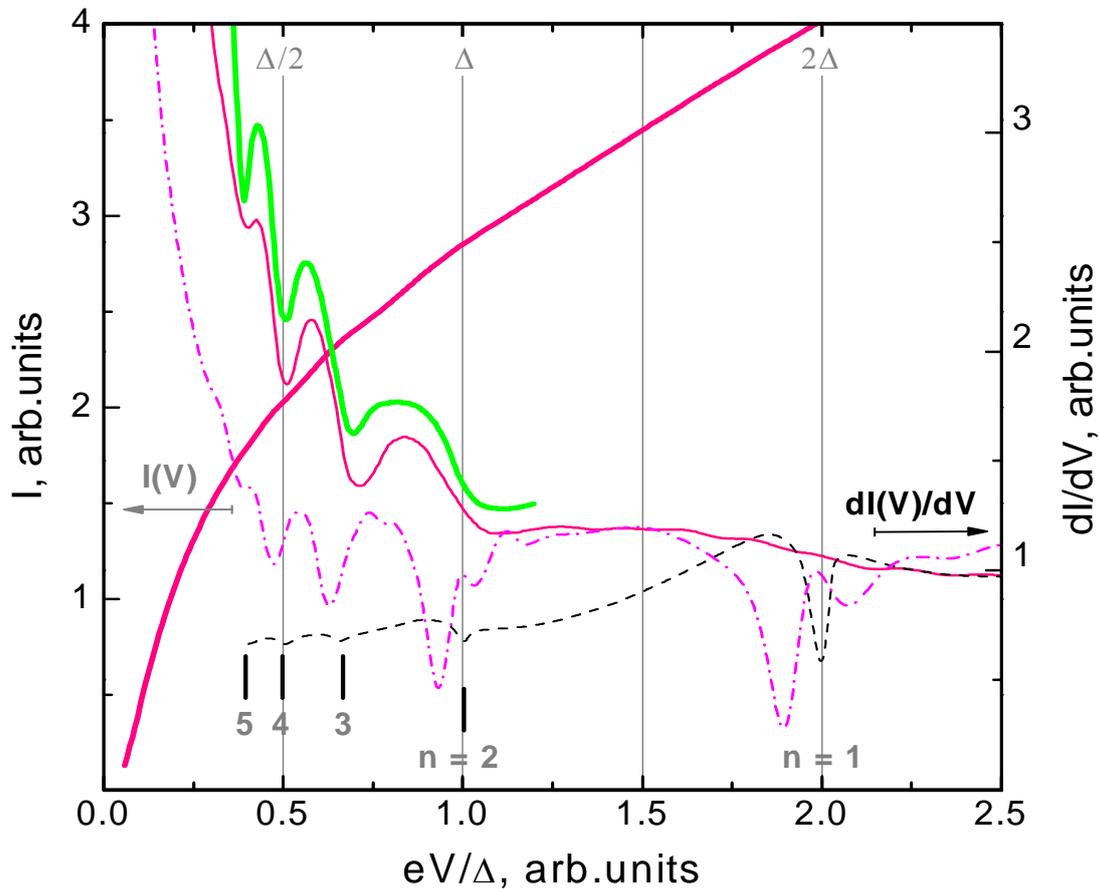

*Fig. 3.* The theoretical dI(V)/dV-spectra (right vertical axis) of SnS contacts, obtained based on the OTBK model (black dashed line, Z=1) [6], Arnold (thick light green line, probability of passing the barrier $T^2=0.83$) [85], Cuevas et al. [87] and Averin–Bardas [86] (thin solid pink line; barrier height h=0.23, transparency 95%; the corresponding CVC is shown by a thick line, left vertical axis), Kümmel et al., dash-dot line; the ratio of the mean free path to the diameter of the contact l/2a=5, T=0.8$T_c$) [88]. The position of the SGS minima is denoted by n=1,2,…

### 3.2. Superconductors with an anisotropic order parameter: Andreev spectroscopy of SnS contacts

Multiple gap superconductors have been widely known since the discovery of $MgB_2$ in 2001 [89], and have been studied intensely since then [54,55,90–95], even though two-gap superconductivity was theoretically predicted in the middle of the last century [96,97].

We will note that two-gap superconductivity was experimentally observed by Ponomarev et al.: they found that the dynamic conductance of tunnel break-junctions in single-crystals Y(Yb)BaCuO could not be described using a single gap model, and reproducibly two series of features were demonstrated [98], which was confirmed by theoretical calculations [99,100].

As is known, the anisotropy of a superconducting order parameter $\Delta$ can be caused by the following factors: basic deviation of the symmetry type from the s-wave, i.e., by the presence of an angular gap amplitude distribution $\Delta(\theta)$ in the k-space [101,102]; the splitting of the gap amplitude, due to degeneracy removing for condensate, realized on the Fermi surfaces with similar geometry and slightly different momentum modules (e.g., nested); by the variation of the gap in the real space, for example due to an inhomogeneous distribution of impurities or dopant. Understanding the causes and consequences of gap anisotropies is extremely important for the determination of HTSC mechanisms [103,104], and therefore interest in this subject has been revived with the study of iron-based superconductors. For example, Ref. [20] contains an analytical investigation of how the gap anisotropy in the k-space affects the shape of the Andreev spectra of NS contacts; unfortunately for SnS Andreev contacts in the c-direction, such detailed theoretical calculations still have not been performed.

We will now consider the process of multiple Andreev reflections and subharmonic structures in dI(V)/dV-spectra of the SnS contact in a multi-gap superconductor, in more detail. A diagram detailing how the carriers are transported across the n-layer of the SnS contact is shown in Fig. 4(a). By applying the bias V, Andreev current will start to flow, in which electrons and holes with any momentum corresponding to the Fermi surfaces of the studied material, will participate. If such an SnS contact is organized in the c-direction (which is exactly the case when using the break-junction) and V is small, then $p_z \ll p_x, p_y$. Since in a ballistic Andreev contact ($l \gg 2a$) the electron momentum is preserved, then the mixing of quasiparticles belonging to different bands does not occur. We can say that each of the bands witnesses the realization of their own Andreev transport channel. Therefore, in the dI(V)/dV-spectrum of an Andreev contact in a multi-band superconductor we should expect the appearance of several SGS, corresponding to each of the gaps.

In the case of a single-gap superconductor having gap anisotropy in the k-space, the shape of the Andreev singularities along the spectrum for such a contact in the c-direction can reflect the anisotropy of the order parameter in the $k_{xy}$-plane (see Fig. 4(a)). Due to the stratification and multiorbital nature of the zones, HTSCs most often have Fermi surfaces that are close to cylindrical, and a characteristic anisotropy of the order parameter $\Delta$ in the $k_{xy}$ momentum space, that corresponds to the ab-plane of the real space, in which its value depends on the direction of the momentum $\Delta = f(k_x, k_y)$. The angle h is usually used in order to consider the type of the order parameter anisotropy, such that $tg(\theta) = k_y/k_x$. Theoretical studies of this problem were initiated rather long ago: for example, in Refs. [101,102], a gap amplitude function that is symmetrical with respect to $k_x$ and $k_y$ was proposed for SIS and NIS contacts based on HTSCs with an anisotropic gap and a van Hove singularity close to $E_F$, $\Delta = \Delta_0 + \Delta_1 \cos(4\pi\theta)$ with four amplitude maxima $\Delta = \Delta_0 + \Delta_1$ in the $\pm k_x$ and $\pm k_y$ directions. We did not want to introduce one more energy parameter ($\Delta 1$), since it does not appear to have any real physical meaning, therefore we will introduce the function $\Delta(\theta)$ in another form: $\Delta(\theta) = \Delta_{max}(1 + 0.5A \cdot [\cos(4\pi\theta) - 1])$, wherein $\Delta_{max}$ is the maximum amplitude, and the coefficient A represents the gap anisotropy in percentages.

Fig. 4(b) shows the collected dynamic conductance curves (in relative coordinates) for highly transparent SnS contacts, with current in the c-direction, which were qualitatively estimated by us using the Devereaux and Fulde [105] calculations for a superconductor with an isotropic order parameter. The background exponential course of all spectra is suppressed, and the calculation is conducted for an "ideal" ballistic contact with $l \gg 2a$ at T=0. In the case of a superconductor with an isotropic gap (pure s-symmetry), the Andreev minima are more intense and symmetrical. Since dI(V)/dV is the sum of partial conductivities of each of the bands, then for a two-gap superconductor with order parameters developed in different bands and having similar amplitudes (we have taken values $\Delta_1=1$ and $\Delta_2=0.9$) SGS will be a series of doublet features (spectrum No. 1 in Fig. 4(b)). It is clear that both minima that make up the doublet are also sufficiently sharp and symmetric, and the conductance reaches background in $eV_n$ intervals between $2\Delta_1/n$ to $2\Delta_2/n$. The spectrum No. 2 shows how the shape of the doublet changes with 10% gap anisotropy in the k-space, i.e., for a distribution that looks like $\Delta(\theta) = 1 + 0.05 \cdot [\cos(4\pi\theta) - 1]$. We can see that the anisotropy in the momentum space leads to the appearance of sufficiently sharp minima having a fine structure, corresponding to the minimum and maximum value of the gap, according to the angular distribution of $\Delta(\theta)$. Both minima are asymmetrical and connected by an arch, but it does not reach the background of the spectrum. The strong anisotropy of the order parameter (Spectra 3–5 in Fig. 4(b)) complicates the interpretation of dI(V)/dV-spectra.

The SGS intensity on spectra 3–5 is highly suppressed relative to the features on Curves 1 and 2 (Fig. 4(b)). Curve 3 corresponds to 50% gap anisotropy in the k-space. Here, in comparison to spectrum 2, the doublet arches become wider: for example, for the main subharmonic with n=1 the minimum corresponding to the lower extremum of the gap merges with the more energetic part of the extremum of the second harmonic $2\Delta_{min}/e = 2\Delta_{max}/2e$; thus visually the spectrum retains SGS that are comprised of asymmetric minima, and the even Andreev minima are more pronounced than the odd. For an arbitrary gap anisotropy that is greater than 50%, the components of the $n^{th}$ Andreev minimum will overlap with features having an of order n+1. It is obvious, when there is 100% anisotropy (i.e., when the order parameter retains the sign, but has nodes in the k-space) there will be a series of asymmetrical features, the position of which will correspond to the maximum amplitude of the gap (spectrum 4). This case is reminiscent of implementing d-wave (sign alternative) gap symmetry (see spectrum 5; obtained in Ref. [105]), with the exception of slightly different shape of the minima and relative amplitudes thereof, which becomes minimal in the case of d-symmetry. In general, unlike the strengthening of the even subharmonics in spectrum 3, in two of the last cases the SGS amplitudes gradually decrease with increases in n.

To summarize, we would like to note that the high quality of the Andreev spectra obtained using the "break-junction" technique allows us to reproducibly observe the fine structure of the Andreev minima and make conclusions concerning the type of gap symmetry. In our experiments it is possible to identify SGS that correspond to gaps with pure s-wave symmetry, and those with anisotropy up to ~40%, as well as gaps with nodes in the angular distribution. It is virtually impossible to distinguish between strong anisotropy, which is greater than ~40%, d-symmetry, and the presence of nodes without changes in the sign of the order parameter (fully anisotropic s-type) using SnS Andreev spectroscopy. In the hypothetical case of the order parameter anisotropy in the c-direction, there will be a broadening and blurring of the Andreev minima with the preservation of both their symmetry and asymmetry.

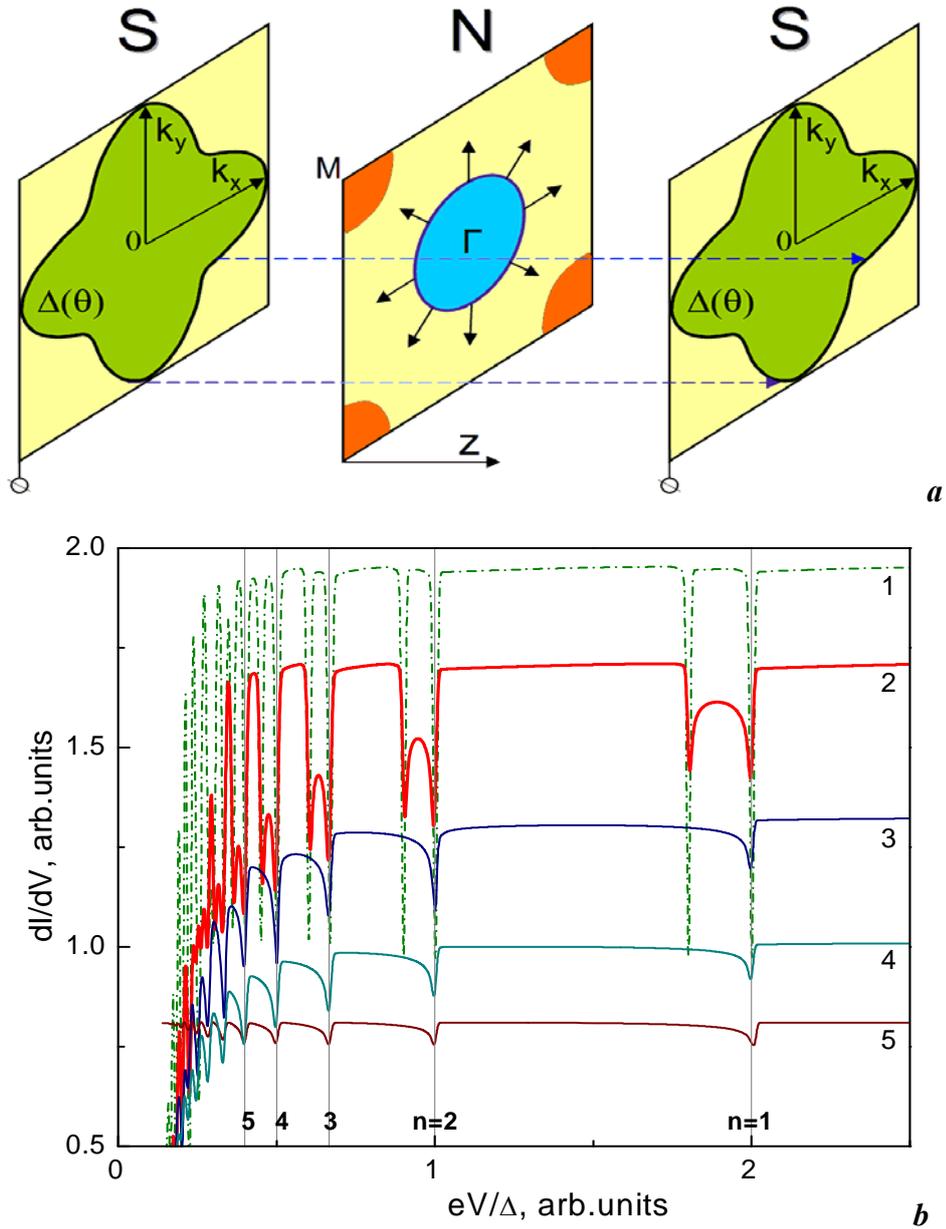

***Fig. 4.*** *A diagram of carrier transport through the SnS contact in the c-direction for the case of a superconductor having an anisotropic gap in the ab-plane. The letter M denotes the center of the electron band (orange), the hole band is arbitrarily denoted by the letter Γ and a blue oval (a). A qualitative calculation of the Andreev features along the dynamic conductance of the SnS contact (based on the results obtained by Deveraux and Fulde in Ref. [105] for a superconductor with an isotropic order parameter): with two independent order parameters with similar amplitudes (Spectrum 1), with 10% gap anisotropy in the $k_{xy}$-plane (extended s-type symmetry) (2), with 50% gap anisotropy in the $k_{xy}$-plane (3), with 100% gap anisotropy in the $k_{xy}$-plane (with nodes, no sign change) (4), with d-wave gap symmetry from Ref. [105] (5) (b).*

## 4. "Break-junction" experiments on layered hightemperature superconductors

*4.1. Tunneling spectroscopy of HTSC cuprates. Intrinsic Josephson effect*

In our experiments on layered HTSCs the constriction can be electrically equivalent to the insulator (SIS contact) as well as to a normal metal (SnS contact). Stratifying the cuprate samples using the technique shown in Fig. 1 is very simple; during this process the microcrack usually separates those blocks of the crystal structure (along the c-direction) that are most weakly bound. Thus, for example, in compounds $Bi_2Sr_2Ca_nCu_{n+1}O_{2n+6}$, two BiO-layers turn out to be separated, located in the middle of the spacer that acts as an insulator. As such, the microcrack creates a contact with a so-called constriction that is electrically equivalent to the SIS tunnel contact. The current-voltage characteristics of the resulting contact usually exhibit a strictly vertical section at V=0, i.e., a Josephson supercurrent, the amplitude of which demonstrates Fraunhofer-type oscillations even in a relatively weak magnetic field according to the law $|sin(x)/x|$ [31], which is the unambiguous proof of the Josephson nature of the supercurrent. The CVCs also exhibit a gap singularity, which is a sharp increase in quasiparticle current at $V=2\Delta/e$, wherein $\Delta$ is the order parameter. At $T \ll T_c$ the value of the superconducting gap can be accurately determined directly from the position of the maximum; as the temperature increases the position of the tunneling peak does not correspond to $2\Delta(T)/e$ and the gap should be determined by approximating the experimental dI(V)/dV-spectrum using the Dynes model [4].

It is experimentally [106–108] and theoretically [109] proven that HTSC cuprates are natural SISI-…-S superlattices in the c-direction: the superconducting $CuO_2$ planes (intercalated with calcium) play the role of the "S", whereas the oxide spacers act as insulator. Therefore, the crystal behaves like a stack (in the c-direction) of series-connected Josephson junctions. The intrinsic Josephson effect [80,110] was first observed in such stack structures, created in Bi-2212: the dI(V)/dV-spectra had tunneling maxima located at multiples of the total gap energy $V=2\Delta \cdot m/e$, wherein m is the natural number of the contacts in the stack. Thus, a unique layered structure of HTSC cuprates makes it possible to study the properties thereof using a method based on the intrinsic Josephson effect, which is intrinsic tunneling spectroscopy. Subsequently, the presence of the intrinsic Josephson effect was confirmed in studies of mesa structures based on HTSC cuprates, by observing CVC branching at a current in the c-direction [106–108], Fraunhofer oscillations of the Josephson mesa structure supercurrent [111] and geometric Fiske resonances [112].

Fig. 5 shows a typical CVC of a tunnel junction (red curve), created in a slightly overdoped Bi-2212 sample with a critical temperature $T_c \approx 88K$ and an order parameter $\Delta \approx 25$ meV, as well as dI/dV-spectra of tunnel SISI-…-S stacks containing m=7 and m=12 contacts (black and magenta curves, respectively). The given CVC refers to the black dI(V)/dV-curve, and the inset on Fig. 5 shows an enlarged fragment of it near zero bias, containing Josephson supercurrent (markedly suppressed by the magnetic field of the Earth). Sharp tunneling maxima corresponding to energies eV = 7·2Δ ≈ 343 meV and eV= 12·2Δ ≈ 596 meV are clearly visible in the dynamic conductance spectra. Note that the position of the main features of the tunneling spectra is symmetric, which excludes the existence of a charge along the surface of the cryogenic clefts. The absence of hysteresis and CVC branching indicates that the SIS contacts that make up these two stacks obtained using the "break-junction" technique, are equivalent. It is obvious that in order to determine the number of contacts in the stack we must normalize the bias voltage axis to the corresponding natural numbers m; after such a normalization the position of the features on the dI(V)/dV-spectra coincides.

It is known that in single crystals of cuprate HTSCs there are screw dislocations, wherein the supercurrent flowing over them shunts the tunneling transport in the c-direction through the SISI-…-S structure. This is why in break-junction experiments on cuprates the step on the cryogenic cleavage that forms the arrays is usually shunted by superconducting banks, a so-called base-to-base contact. The parallel base-to-base contact also makes a significant contribution do the dynamic conductance of the break junction in the form of tunneling peaks at the bias voltages |V| = 2Δ/e and therefore, corresponds to a single SIS contact. Once again we will note that the use of the current source and hardware recording of the dI/dV and not dV/dI dependence guarantees the constancy of the scale for the bias V in the case of contacts being formed in parallel to the studied one.

Similar features of the shunting base-to-base contact at eV ≈ 52 meV are clearly visible on dI(V)/dV-spectra of the stacks shown in Fig. 5 and are marked by gray arrows. Note that the position of the peaks of the quasiparticle tunneling current for the base-to-base contact is not scaled with the number of contacts m in the stack under study and does not change during the process of mechanical readjustment, in contrast to the main gap features of a stack contact. First of all, this suggests that the peaks at 53 mV cannot be interpreted as gap singularities from the second order parameter (because in that case their position would also scale in sync with m). Secondly, these singularities are the manifestation of the bulk superconducting gap; during the measurement of the dI(V)/dV-spectrum over 0 < T < $T_c$, the maxima of the base-to-base contact define the temperature dependence of the gap, corresponding to the data for the stack contact. Consequently, the impact of the base-to-base contact can always be uniquely identified in the experiment. At the same time, the observation of features related to the SIS base-to-base contact allows us to directly obtain the value of 2Δ, which significantly simplifies the determination of the number of contacts in the studied stack.

During the process of mechanical readjustment the cryogenic surfaces slide along the ab-planes: the touch point "hops" from one terrace to another; the height of the steps changes, as does the number of "acting" layers and therefore, so does the number of contacts in the stack m. If the bend of the substrate is varied with precision, we are able to get SISI-…-S structures with different m; in this case the position of the tunneling peaks along dI(V)/dV-spectra will change in multiples of 2Δ. In the study authored by Ponomarev et al. [30,43], arrays consisting of m = 2–25 contacts were observed on the same sample; a typical resistance R = 2–2000Ω per SIS contact allowed us to record I(V) and dI(V)/dV-characteristics of the stacks with a large number of contacts, virtually eliminating overheating.

In Fig. 5 the features of the stack tunnel junction (m=12) and of the single base-to-base contact have been fitted for the magenta curve as an example. An approximation according to the Dynes [4] model does not take into account the anisotropy of the superconducting gap in the momentum space Δ(θ), but nevertheless, it allows for a satisfactory description of both the tunnel features and to estimate the broadening parameter Γ. Since for a stack contact Γ = 2%–4% of energy Δ, whereas for a base-to-base contact Γ ≈ 25%, which originates from defectiveness of what seems like an ideal cryogenic surface, we have reason to conclude that natural arrays are much more reliable objects for the study of the physical properties of layered superconductors in comparison to the surface of cryogenic cleavages.

In our experiments the values of the order parameter and their temperature dependences that are obtained using Josephson spectroscopy of single SIS contacts and intrinsic Josephson spectroscopy of SISI-…-S structures coincide, are reproducible, and do not depend on the dimension and resistance of the contacts, or on the number of contacts in the stack.

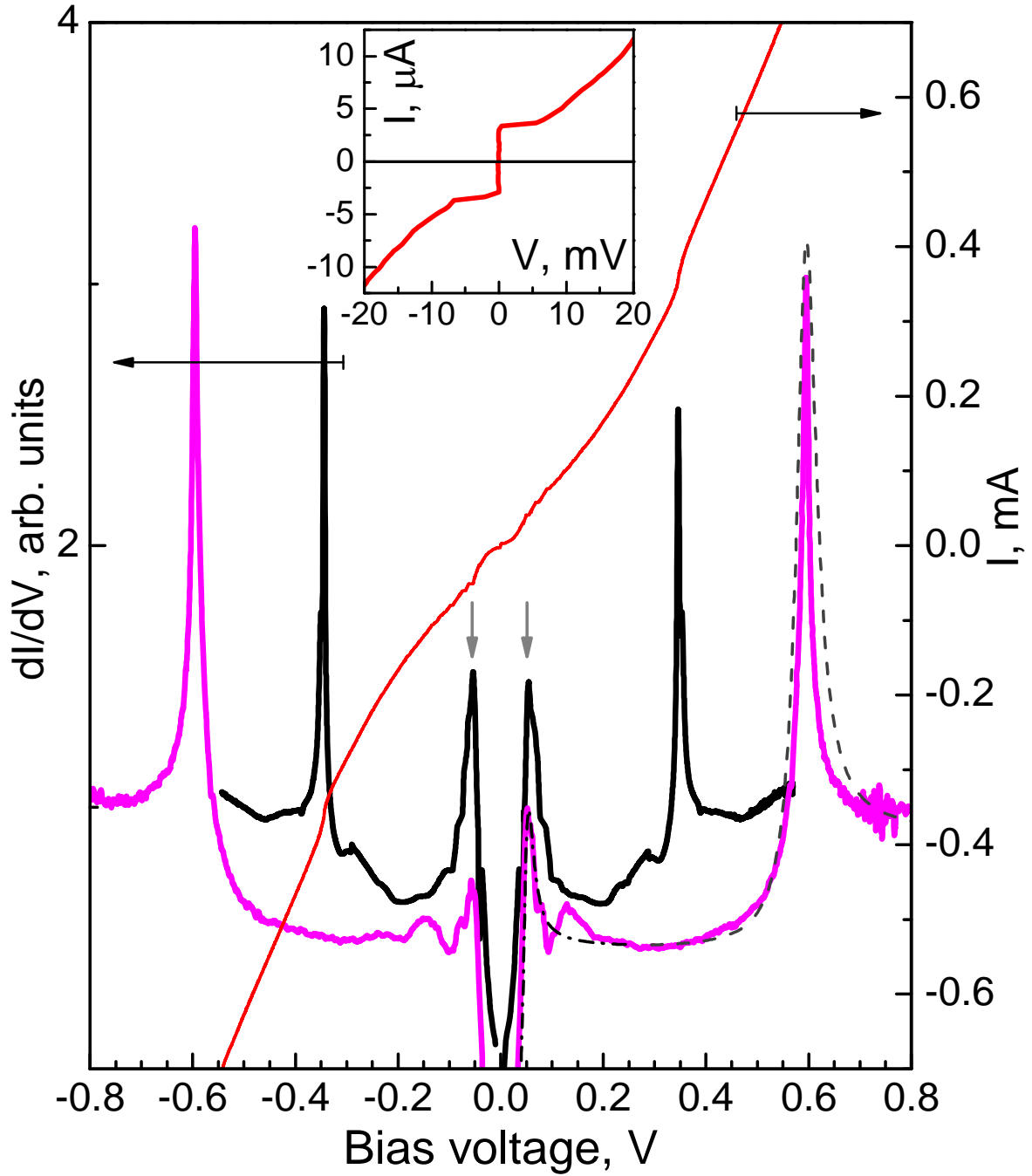

***Fig. 5.*** *Dynamic conductance spectra dI(V)/dV (left axis) of stack tunnel SISI-...-S structures (black and magenta curves correspond to the arrays consisting of 7 and 12 consecutive junctions) in overdoped Bi-2212 with $T_c \approx 88$ K. T = 4.2 K. The current-voltage characteristics (red curve referring to the right vertical axis) are given for the stack made of 7 junctions (black dI(V)/dV-dependence). The gray arrows indicate the tunneling peaks that correspond to $2\Delta/e$ in single SIS base-to-base contacts. The inset depicts the enlarged CVC fragment that shows the Josephson supercurrent. The dashed and dash-doted lines correspond to fragments of theoretical dI(V)/dV, calculated based on the Dynes model [4]. Adapted from Ref. [119].*

*4.2. The break-junction method as an instrument of studying the optical phonons in HTSC cuprates*

The high quality of the CVC and dI(V)/dV-characteristics of the Josephson junctions and stack structures obtained via "break-junction" experiments has provided the opportunity to resolve not only the main gap singularities, but also the fine structure of the spectra. In Refs. [8,30,36,37], a clear reproducible structure that corresponds to the excitation of optical phonons was observed. Fig. 6 (Ref. [113]) can serve as example of obtaining the fine structure that corresponds to the energy of optical Raman-active phonon modes. For underdoped Bi-2212 samples with $T_c = 21$ K, a set of features at biases $V < 2\Delta/e$ is clearly visible. The fact that these resonances are caused specifically by the alternating Josephson supercurrent is easy to verify: as the amplitude of the latter is suppressed by a relatively weak magnetic field, the amplitude of the features along the dI(V)/dV-spectrum also decreases. The experiment shows that changes to these amplitudes are proportional, and with complete suppression of the Josephson supercurrent the features that can be associated with optical phonons, disappear. A reliable test is also the variation of the temperature: in Fig. 6 and the corresponding temperature dependence of the peculiarities in Fig. 7(a) we can clearly see that the position of the fine structure is not affected by this variation. The phonon energy is determined by the position of the fine structure features such as $2eV_{ph}$ and corresponds to the position of the optical phonon modes in HTSC cuprates, which are determined according to Raman studies [114–117].

Ponomarev et al. have shown [8,30,37] that the resulting phonon frequencies do not depend on the temperature (see Figs. 6 and 7(a)), the dopant concentration (see Fig. 7(b)), and the number of $CuO_2$ planes, and are reproducible via different families of cuprate samples. Thus, we conclude that in SIS contacts obtained in the c-direction of the crystal lattice, the AC Josephson supercurrent resonantly excites the coherent Raman-active optical phonons, which points to the importance of a strong electron-phonon interaction in the high temperature superconductivity mechanism of cuprates [30,36,37,118]. Note also that we did not discover any resonant excitation of magnons in SIS break-junctions [8,30,37].

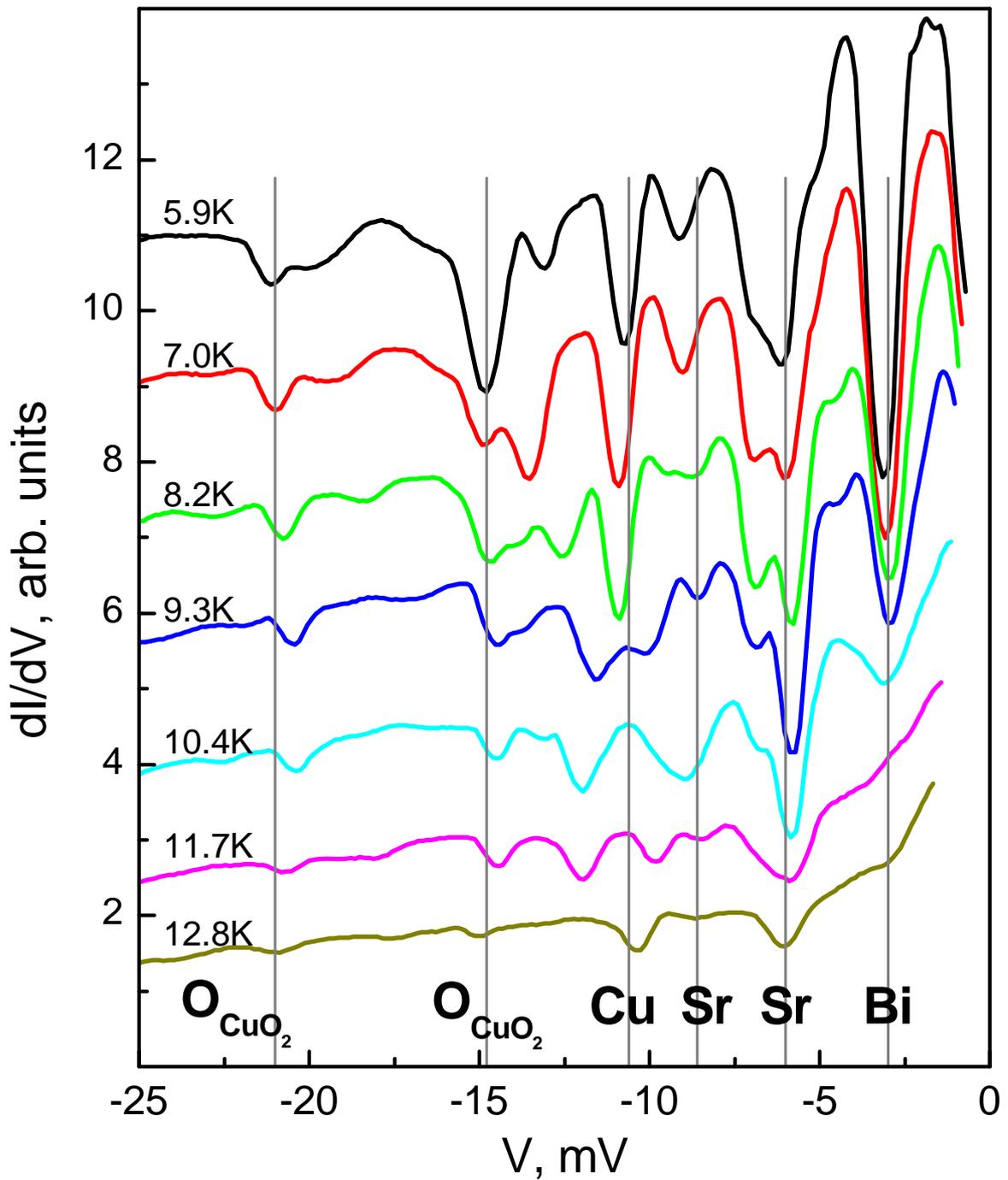

*Fig. 6. A fragment of the dynamic conductance spectrum of a SIS break-junction in an underdoped Bi-2201(La) sample with $T_c \approx 21K$ within the 5.9–12.8K temperature range. The position of the singularities caused by the interaction of the AC Josephson current with the optical phonon modes is marked by vertical lines. Derived from Ref. [113].*

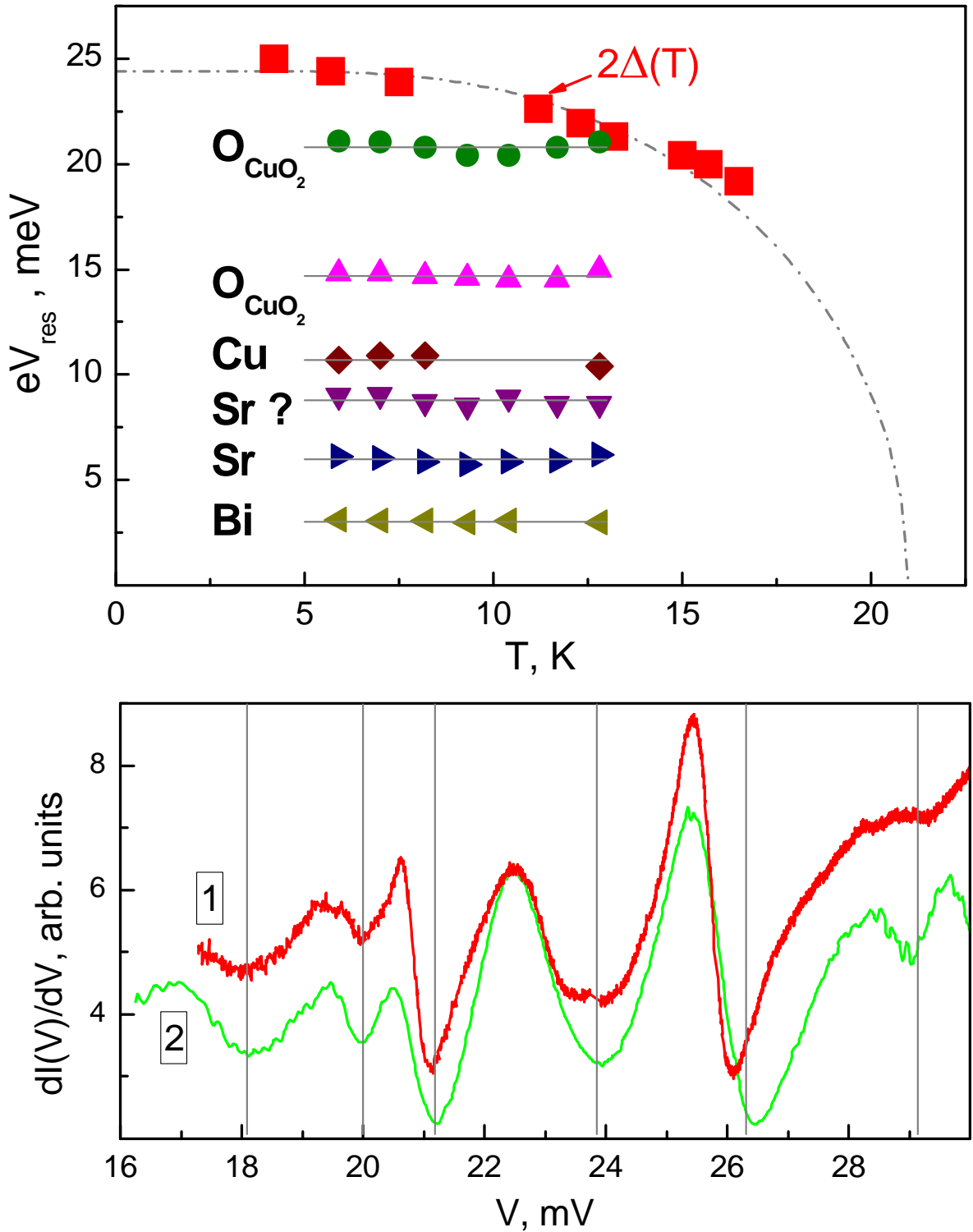

*Fig. 7. The temperature dependence of the superconducting gap 2Δ(T) (■) and the positions of the features along the dI(V)/dV-spectrum, caused by the interaction of the AC Josephson current with the optical phonon modes (circles, triangles, diamonds) (according to data from Fig. 6) (a). A comparison of the fine structure of a portion of the dI(V)/dV-spectra (T = 4.2 K) of SIS contacts obtained in underdoped (1) and optimally doped (2) Bi-2212(La). The positions of the phonon resonances are reproduced and shown by vertical lines (b). Derived from Ref. [113].*

*4.3. The manifestation of the effect of multiple Andreev reflections in the experiment*

In samples BiSr(La)CaCuO [30,31,81] the lanthanum dopant atoms changed the structure of the spacer, as a result of which the constriction region of the break-junction often exhibited properties of a thin (ballistic) normal metal layer. In the experiment we observed CVC and dI(V)/dV-spectra that are typical for pure SnS Andreev regime with a high transparency [6,85–88]. Later the Andreev contacts were obtained in YBaCuO [29,44,45].

If under certain conditions the spacer formally exhibits properties of a normal metal, then by analogy with the intrinsic Josephson effect we can assume that the layered structure of the crystal is implemented as an SnSn-…-S stack of equivalent Andreev contacts. Indeed, the group led by Ponomarev first observed the effect of intrinsic multiple Andreev reflections82 by making the break-junction on BiSr(La)CaCuO samples. The position of the Andreev subharmonics on the dynamic conductance spectrum was scaled by an integer multiple compared to dI(V)/dV-single SnS contact:

$$V_n = 2\Delta \times m/en, \qquad (2)$$

wherein m is the number of contacts in the stack; m, n are the natural numbers.

Subsequently the intrinsic multiple Andreev reflections effect is repeatedly observed in other layered materials: cuprates of different families [30,36,37,119], $Mg_{1-x}Al_xB_2$ [36,48,49,51,52], both single crystal and polycrystalline iron-based superconductors [66–69,71,72,74–77]. Note that the features created by the contribution of base-to-base contacts in the SnS Andreev regime were observed very rarely. This could be associated with the fact that a large size of base-to-base contacts (comparable to the width of the terraces) at times exceeds the mean free path of the carriers, i.e., it does not ensure ballistic transport and prevents the observation of Andreev reflection.

A large number of observed Andreev subharmonics (up to n=5) in cuprate samples allowed us to accurately determine the amplitude of the gap, while obtaining both the tunnel and Andreev dI(V)/dV-spectra on one sample allowed us to collect data statistics on four methods and provided reproducibility. After normalizing the spectra to the corresponding number of contracts m in the stack, the position of the main gap features, which were a maximum in the tunnel regime and a minimum in the Andreev regime, coincided at the 2Δ/e bias. Thus, four methods of tunneling spectroscopy (two surface and two bulk) implemented using the "break-junction," demonstrated the identical values of the superconducting gaps, thereby confirming the bulk nature of the observed order parameters Δ and the accuracy of the obtained results [36,37]. The scaling Δ and $T_c$ was observed for both under and overdoped cuprates based on Bi, Hg and Tl [13,30,36,37,119]. As such, in "break-junction" experiments we probed namely the superconducting order parameter, and not the pseudogap.

*4.4. Tunneling and Andreev spectroscopy of $Mg_{1-x}Al_xB_2$*

Magnesium diboride, having a layered crystal structure and the highest critical temperature (in a stoichiometric state and under the normal pressure) for binary compounds $T_c \approx 40$ K [89], is in many ways similar to HTSC cuprates [36,37,120]. However, since the strong isotope effect of boron clearly points to the phonon nature of the pairing mechanism in $MgB_2$ [121] then for many researchers the desire to describe the superconductivity phenomenon in $MgB_2$ only using the strong electron-phonon interaction and the values of the characteristic ratio $2\Delta/k_BT_c < 5$, is especially attractive [122–124]. As we already know, the most surprising result of the theoretical studies involving magnesium diboride was the prediction of two-gap superconductivity [125,126]. Two types of bands cross the Fermi level of $MgB_2$ (quasi-2D hole σ-band close to the Γ point of the Brillouin zone, formed by the s-orbitals of boron, and three-dimensional electron and hole π-bands, formed by the p-orbitals of magnesium close to the M point), in which at $T<T_c$ at least two independent superconducting condensates are developed. Reference [126] accounted for the splitting of the Fermi surface sheets in the σ- and π-bands that caused the appearance of doublet peaks in the theoretical quasiparticle density of states and the implementation of a four-gap approximation (i.e., the presence of two close σ-gaps and two almost identical π-gaps).

In our experiments the crystal lattice of the $Mg_{1-x}Al_xB_2$ samples formally exhibited both insulator and normal metal properties, and SIS and SnS stacks formed on the steps and terraces of the cryogenic clefts due to the layered structure [36,37,48–62]. As such, we got the opportunity to apply four types of spectroscopy to the study of the structure of superconducting order parameters, just as we did with cuprates.

The break-junction spectra we obtained for $Mg_{1-x}Al_xB_2$ are not describable by the single-gap model. The dI(V)/dV-spectra of the junctions in MgB2 samples with a maximum $T_c \approx 40$K had gap singularities from a large gap $\Delta_\sigma = (10.5 \pm 1)$ meV and a small gap $\Delta_\pi = (2 \pm 0.5)$ meV. We observed scaling of the σ-gap and the $T_c$ within $T_c = 11–40$K during both increases in the concentration of aluminum dopant and increases in the degree of disorder of the crystal structure; the amplitude of the π-gap remained nearly constant under these conditions, all the way to $T_c \approx 15$K (which corresponds approximately to the "eigen" critical temperature of a 3D π-condensate in the hypothetical case of a zero interband interaction), after which at $T_c < 15$K it decreased monotonically. We did not observe the isotropization of the order parameters (transition to the "dirty limit") until $T_c \approx 11$ K [36,48,51], which was expected in Refs. [127,128].

Fig. 8 shows fragments of the dI(V)/dV-spectra of break junctions, containing singularities caused by a π-gap and obtained in a polycrystalline sample $MgB_2$ with $T_c \approx 35$K using a sequential mechanical readjustment [113]. The bottom dI(V)/dV-spectrum in Fig. 8 corresponds to the SnS Andreev junction with a high transparency: the first and second Andreev subharmonics are clearly visible at the biases $V_1 \approx \pm 4$ mV and $V_2 \approx \pm 2$ mV, respectively. Therefore, according to Eq. (1), the amplitude of the small gap is $\Delta_\pi \approx 2$ meV. At large bias voltages we also see an Andreev minimum ($n_\sigma = 2$), the position of which corresponds to the large gap $\Delta_\sigma \approx 8$ meV. During gentle mechanical readjustment of the contact geometry the main $2\Delta_\pi$-minima turned into maxima: the top spectrum of the dynamic conductance is typical for SIS contact, and at the same time the position of the tunnel peaks and the amplitude of the π-gap did not change. We can assume, when the cryogenic surfaces sliding along the ab-plane the transparency of the ScS contact constriction decreased, which, in accordance with the models in Refs. [85,86] caused a transition from the Andreev regime (with an excess current) into the tunnel mode (with a lack of current). Such a SnS to SIS transition of the same contact has been repeatedly observed in break-junction experiments on $Mg_{1-x}Al_xB_2$ (Refs. [48,49]) and cuprates [29,44,45,119]. Presumably, this effect was observed for the first time in niobium break junctions by Müller et al. [23].

In $Mg_{1-x}Al_xB_2$ polycrystalline samples we also revealed the effect of intrinsic multiple Andreev reflections (IMARE) [36,48,49,51]. Fig. 9 shows the CVC (left vertical axis) and dynamic conductance spectra (right axis) for two arrays obtained using $MgB_2$ samples (from the same batch) with critical temperatures of $T_c \approx 40$K. A CVC with noticeable excess current at low bias ("foot") is typical for the IMARE in a ballistic SnS contact (high transparent). It is possible to use the product of the bulk resistivity and the mean free path of the carriers $\rho l \approx 2 \times 10^{12}$ $\Omega cm^2$ [129] as well as the Sharvin formula [84] $R = 4\rho l/(3\pi a^2)$ to estimate the diameter of the contact 2a. For the $MgB_2$ polycrystalline samples studied by us, with a maximal $T_c$, $\rho \approx 2$ μΩ·cm, it is nevertheless demonstrated that the resistivity of a single crystal is at least four times lower [130] then $l \approx 40$ nm; it is these latter values of $\rho$ and $l$ that are relevant to us, because break-junction is created within diameter d of the split grain. When we set the typical resistance of the SnS contacts obtained by us to R ≈ 1–60Ω (see Figs. 9 and 10), we get 2a ≈ 2–18 nm << $l$ < d. The resulting diameter turned out to be less than the mean free path of the carriers, which points to ballistic transport across the break-junction, thus giving us the option of studying the effect of multiple Andreev reflections. From the assessment above it follows that the value 2a is orders of magnitude less than the size of the grains in the $MgB_2$ polycrystalline samples we used [131], which is consistent with our assumption about the local formation of stack nanostructures on the steps and terraces of the cleaved crystallites.

The CVCs and dynamic conductance spectra (see Fig. 9) of the array were normalized to m=5 (top curve) and m=2 (bottom curve). The number of contacts in the stack was determined by choosing the smallest natural number, by which the bias axis could be divided, in order to align the positions of the gap singularities. On the bottom dI(V)/dV characteristics we can easily spot the two-gap SGS. The intense minima located at $eV_{n\sigma=1} \approx 19.2$ meV, $eV_{n\sigma=2} \approx 10$ meV and $eV_{n\sigma=3} \approx 6.9$ meV and labeled in Fig. 9 by vertical solid lines, in accordance with Eq. (1), satisfy the positions of the first, second and third subharmonics of the r-gap Dr⏐ 10 meV with the characteristic ratio of the BCS theory $2\Delta_\sigma/k_BT_c \approx 5.8$. Note that in the attempt to interpret these spectra as the corresponding 10- and 4-contact stacks, the SGS formula would give a value for the large gap that is $\Delta_\sigma \approx 5$ meV and a characteristic ratio $2\Delta_\sigma/k_BT_c \approx 2.9$, which is below the weak-coupling BCS limit 3.52 and would be impossible for the leading band. The minima in the dI(V)/dV-spectra that are located at smaller biases $eV_{n\pi=1} \approx 3.8$ meV and $eV_{n\pi=2} \approx 1.9$ meV and denoted by vertical dotted lines and arrows in Fig. 9, have much greater intensity relative to the Andreev singularities with n>2 from the large σ-gap, do not correspond to the position of its fourth subharmonic (which according to Eq. (1) is expected at $eV_{n\sigma=4} \approx 5$ meV) and therefore, cannot be attributed to SGS from $\Delta_\sigma$. The given minima are, obviously, the SGS from the small gap $\Delta_\pi \approx 1.9$ meV. We can see that the obtained values of σ and π-gaps are reproducible and do not depend on the number of contacts in the stack which clearly shows their bulk nature.

Figure 10 shows the CVC (left vertical axis) and dynamic conductance spectra (right axis) for two 4-contact SnSn-…-S stacks, obtained for the same $MgB_2$ sample with a critical temperature of $T_c \approx 40K$ (curves on Fig. 10 normalized to the bias axis at m=4). The contacts were formed sequentially by gentle mechanical readjustment, in the process of which, regardless of the variation in contact dimensions, there was practically no change in the shape of the dynamic conductance spectrum, or the position and shape of the Andreev SGS minima from Dr and $\Delta_\pi$. Therefore, the observed features cannot be attributed to any geometric resonance. Moreover, we can be certain of the fact that the slight change to the contact resistance (see Fig. 10) was caused by a change to the contact area between two cryogenic clefts, sliding along the same terrace.

The validity of the gap amplitudes determined by us in the $Mg_{1-x}Al_xB_2$ system is confirmed by the consistency between these four types of spectroscopy, implemented by the "break-junction" technique. In particular, obtaining similar values for the small gap in experiments on arrays favors of the bulk nature of the order parameter $\Delta_\pi$ and the impossibility of interpreting the observed features as manifestation of the surface gap. Note that the maximum values of the large gap $\Delta_\sigma$ = 10–11.5 meV that were determined for $MgB_2$ by us using break-junctions [36,37,48–53] and by the group led by Ekino $\Delta_\sigma$ = 9–12 meV [34,37,48–53], coincide. The values of $\Delta_\sigma$ were confirmed by that same group using point contact spectroscopy (PCAR) [94]. However, other research teams using STM and PCAR techniques obtained smaller values of $\Delta_\sigma$ = 6–8 meV (for review see Refs. [20,90,91]), demonstrating the significant sensitivity the superconducting properties of magnesium diboride have to the surface quality.

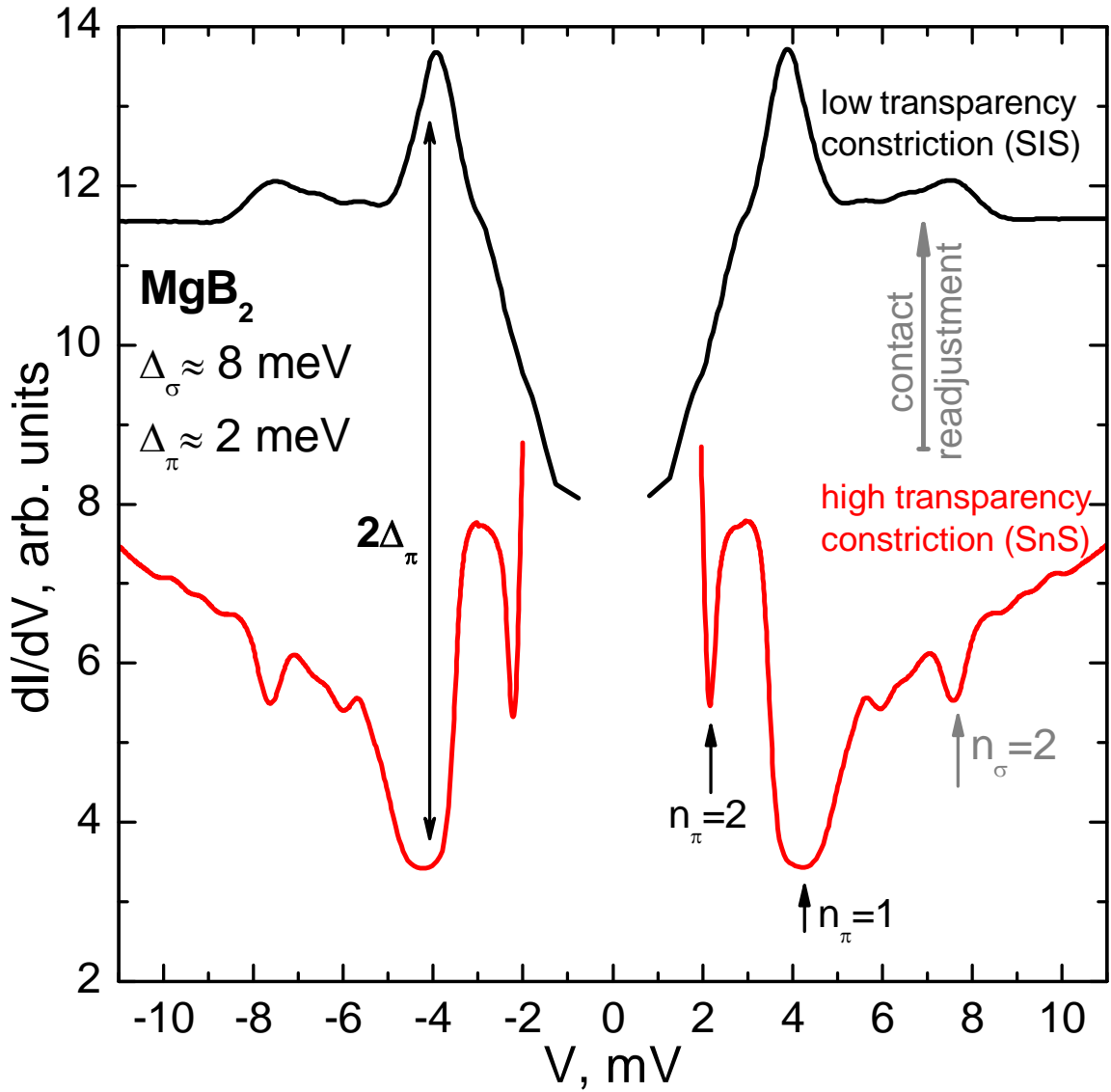

*Fig. 8.* *The transition of the ScS contact constriction in MgB$_2$ with T$_c$ ≈ 35K from a high transparency mode (SnS, bottom dI(V)/dV-spectrum) to the low transparency mode (SIS, top spectrum) for π-band carriers. T=4.2 K. The position of the SGS minima for $\Delta_\pi$ ≈ 2 meV is shown by black arrows and labels n$_p$=1,2; the tunnel peaks of the p-gap are labeled as 2$\Delta_\pi$; the Andreev minimum nr¼2 from a large gap $\Delta_\sigma$ ≈ 8 meV is denoted by a gray arrow.*

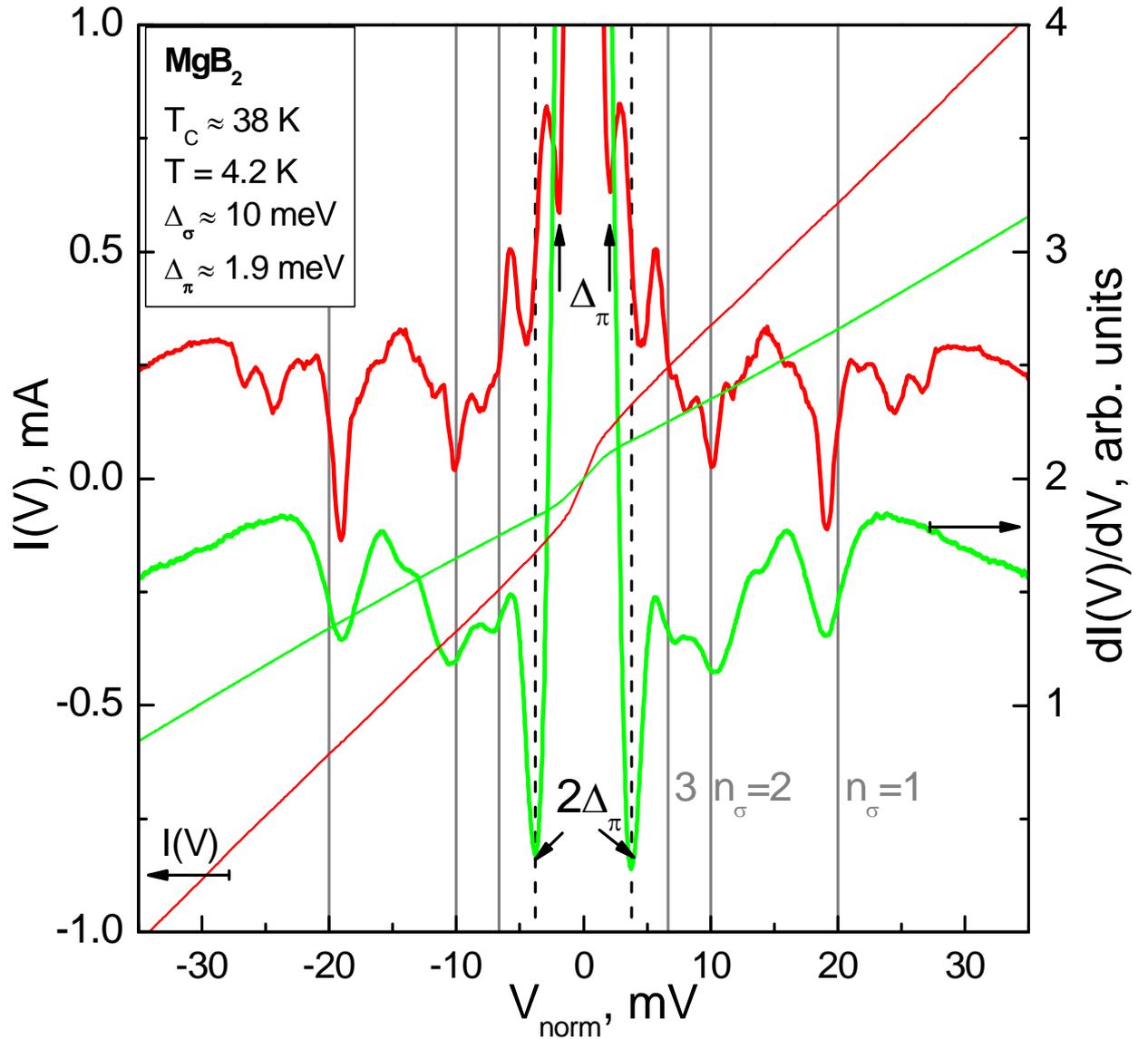

*Fig. 9.* CVC normalized to that of a single-contact (left vertical axis) and dI(V)/dV-spectra (right axis) of Andreev SnSn-...-S structures (top spectrum is m=5 contacts, bottom spectrum is m=½2), obtained for MgB$_2$ samples from the same batch with T$_c$ ≈ 40 K. T=4.2 K. The SGS minima from the large gap $\Delta_\sigma$ ≈ 10 meV are marked by gray vertical lines and the labels n$_s$=1,2,3; SGS from the small gap $\Delta_\pi$ ≈ 1.9 meV are marked by dotted lines and arrows.

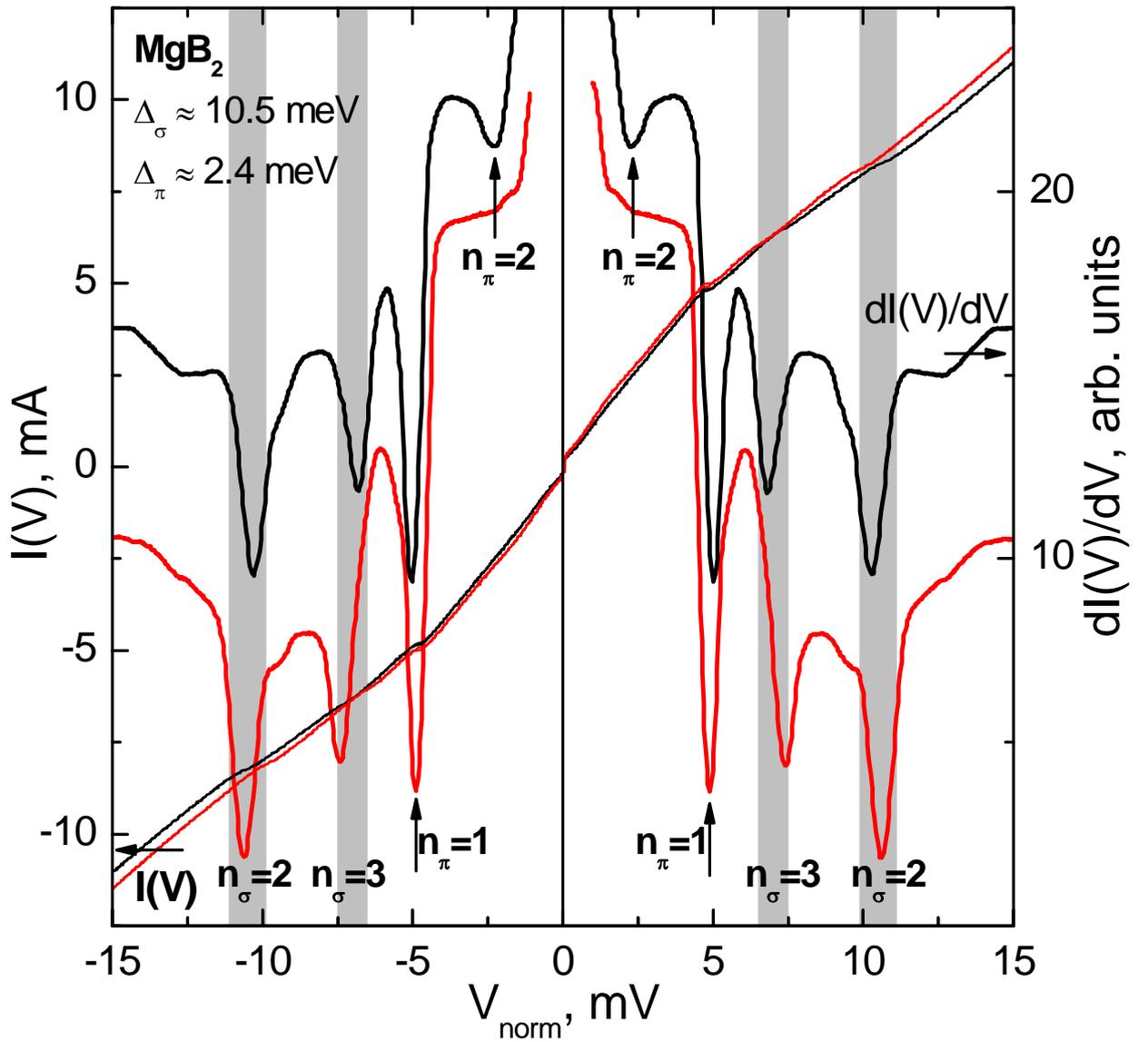

*Fig. 10.* CVC normalized to that of a single-contact (left vertical axis) and dI(V)/dV-spectra (right vertical axis) of Andreev SnSn-...-S structures (m=4 contacts), obtained sequentially by mechanical readjustment on the same $MgB_2$ sample with $T_c \approx 40K$. T=4.2K. SGS minima from the large gap $\Delta_\sigma \approx 10.5 meV$ are marked by gray vertical lines and labels n¼2, 3; SGS from the large gap $\Delta_\pi \approx 2.4$ meV are marked by arrows and labels $n_\pi$=1,2.

### 4.5. Selective transparency mode of ScS contacts based on $Mg_{1-x}Al_xB_2$

In the process of studying ScS contacts based on $Mg_{1-x}Al_xB_2$, we repeatedly obtained dynamic conductance characteristics on which the structure from the large gap corresponded to the high transparency regime of the constriction, whereas the features from the small gap corresponded to low transparency (Fig. 11). As such, though formally it is a metal for σ-holes, a constriction can serve as a thin insulator for Cooper pairs from the π-bands. So, what is it that determines the transparency of the ScS contact based on $MgB_2$? Generally speaking, by changing the distance between the superconducting banks, it is possible to vary the transparency of the barrier in the experiment [82]. Alternately, BTK theory [5] developed for NIS contacts, indicates that the transparency parameter, usually denoted as Z, can be defined by the ratio of Fermi velocities in the superconductor and the metal: $Z = (1 - v_F^S / v_F^N) / 2\sqrt{v_F^S / v_F^N}$. We can assume that since in $MgB_2$ the value of the Fermi velocity $v_F$ in the c-direction for the Cooper pairs from the π-bands (according to calculations in Ref. [132]) is about 9 times greater than the $v_F$ for the σ-pairs, then the ratio of these velocities to the $v_F$ of the thin metal interlayer (the role of which can be played by Mg layers with a destroyed ordering of atoms, located on the surface of the cryogenic cleavage) will also vary approximately by an order of magnitude. This provides the difference between the tunnel barrier transparency parameters $Z_{\sigma,\pi}$ for carriers from the σ and π-condensates, respectively, and also for the latter the transparency turns out to be 2–4 times lower than for the carriers from the σ-condensate. Therefore, for p-pairs the tunneling current through the ScS contact is more probable, whereas for σ-carriers Andreev transport is more likely (the constriction behaves like normal metal). The existence of this regime of selective transparency is indirectly described in Ref. [133].

Fig. 11 shows the CVC and the dynamic conductance of ScS contacts with a selective transparency, created in a $MgB_2$ sample with $T_c \approx 40$ K. The positions of the four observed minima of the subharmonic structure from the rgap $\Delta_\sigma \approx 10$ meV are noted by the labels $n_\sigma=1,2,3,4$. The doublet nature of the minima should be highlighted; we can assume that this is caused by the splitting of the order parameter in the σ-bands (15%–20%): by the degeneracy removing of σ-condensate that open on nested hole Fermi surface cylinders close to the Γ point (which is consistent with the theoretical predictions of the four band model from Choi at al. [126]). At the same time, such doublets for $\Delta_\pi$ were not observed in our experiment, which is possibly associated with the isotropization of the order parameter in 3D π-bands. The inset of Fig. 11 shows the dependence of the Andreev subharmonics position (we take the doublet midpoint) $V_{n\sigma}$ for the σ-gap on their inverse serial number. In accordance to the predictions of Eq. (1), we get a linear dependence that goes through the origin. The Josephson supercurrent on the CVC of this contact and the gap maxima on eV = 2Δ$_\pi$ ≈ 3.8 meV speak in favor of implementing the Josephson effect for carriers in the π-band. Such type of dI(V)/dV-spectra is reproducible by mechanical readjustment of the break junction and does not depend on the geometric dimensions of the contact (Fig. 12).

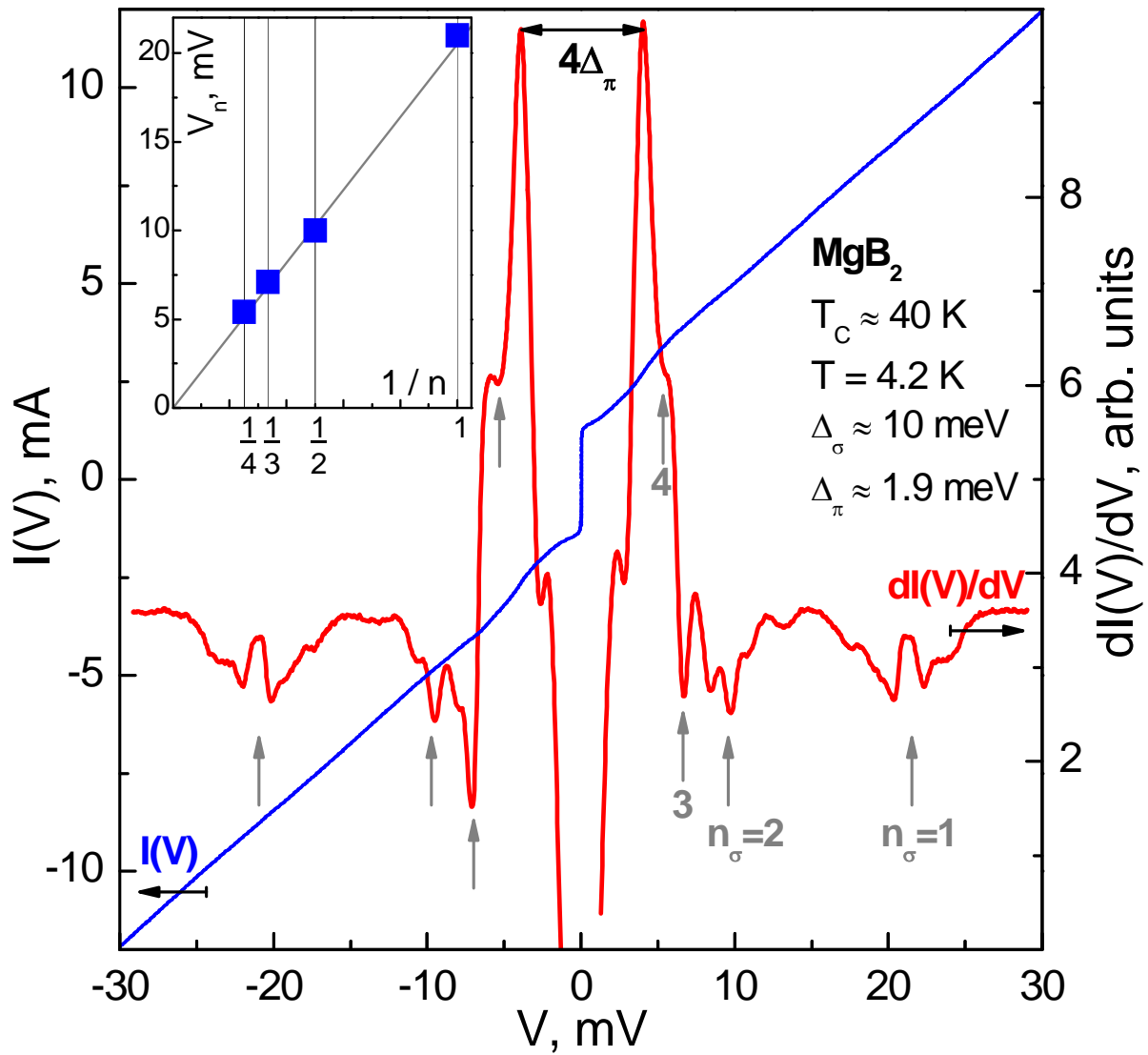

*Fig. 11. The CVC (left vertical axis) and dynamic conductance spectra (right axis) of a contact in selective transparency mode, obtained in MgB$_2$ sample with T$_c$ ≈ 40 K.113 T=4.2 K. SGS from the large gap Δ$_\sigma$ ≈ 10 meV are denoted by gray arrows and labels n$_\sigma$=1,2... (the inset shows the dependence of the SGS doublet midpoint position on their inverse number 1/n for Δ$_\sigma$) The constriction has a high transparency for 2D σ-carriers, however for a 3D π-band the transparency is low. Tunnel maxima from the small gap Δ$_\pi$ ≈ 1.9 meV are shown by black arrows. Adapted from Ref. [113].*

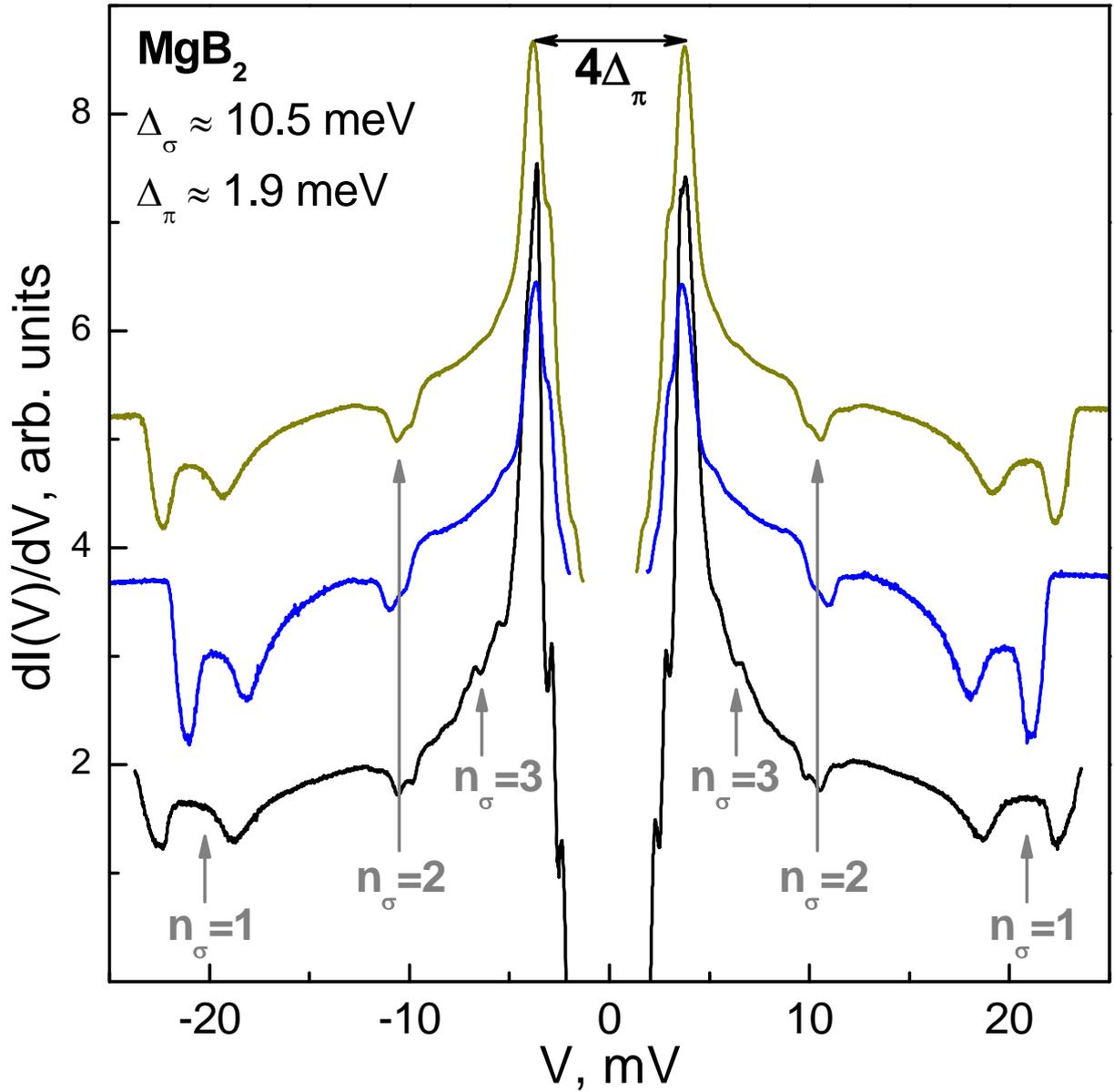

*Fig. 12. dI(V)/dV-spectra of ScS contacts in the selective transparency mode, obtained from the MgB$_2$ sample with $T_c \approx 40K$ by sequential mechanical readjustment [113]. T=4.2K. The constriction has a high transparency for 2D σ-holes, and a low transparency for 3D π-carriers. The Andreev subharmonics of the σ-gap $\Delta_\sigma \approx 10.5$ meV are denoted by gray arrows and labels $n_s$=1,2,3. Tunnel maxima from the small gap $\Delta_\pi \approx 1.9$ meV are shown by black arrows.*

*4.6. (Intrinsic) Andreev spectroscopy of iron pnictides and chalcogenides*

The discovery of high-temperature superconductivity in iron pnictides and chalcogenides [134] marked the beginning of a new stage in intensive research of HTSCs. The superconductivity mechanism in these metals is still unclear. At the moment there is no doubt in the fact that due to the multiorbital nature of the electron and hole bands in these new materials, at $T < T_c$ several order parameters with expressed anisotropy [103,104] are realized. Two mechanisms behind the formation of Cooper pairs are proposed. The so called $s^{\pm}$-model [104,135], based on the proximity of the antiferromagnetic order and the strong influence of the spin fluctuations, predicts certain principles for the developing of a sign-alternating order parameter in iron-based superconductors from different families [136]; yet the recent discovery of nonmagnetic ThNFeAs with a critical temperature of $T_c \approx 30K$ (Refs. [137,138]) does not fit into the developed theoretical framework. An alternative $s^{++}$-model does not deny the importance of accounting for spin fluctuations, and also considers pairing by orbital fluctuations, enhanced by phonons, as the main mechanism. This model successfully describes the experimental data, and the anisotropy of the order parameter is explained by the competition of the spin and orbital interactions [103,139].

The existing experimental data on the amplitude and anisotropy of the order parameter in iron-based superconductors are highly contradictory, although the importance of accurately determining of the gap structure is obvious in establishing the mechanism of superconductivity in iron pnictides and chalcogenides.

In our "break-junction" experiments on iron superconductors from different families, SnS Andreev high transparency regime was implemented. The tunneling spectra were observed only in selenide single crystals $KFe_2Se_2$ with $T_c \approx 18K$ (Fig. 13). We would like to highlight that the "break-junction" technique seemed like one of the few methods that could probe the gap structure of 122-selenide [73,140]; at any degree of purity and stoichiometry, this class of compounds tends to undergo phase separation, but a single superconducting phase is formed. At the same time, the tunnel ScS break-junction can be formed specifically between superconducting regions and is not affected by the influence of non-superconducting phases. Fig. 13 shows the CVC and the dynamic conductance spectrum (T=4.2K) of a stack of m=3 SIS contacts based on $KFe_2Se_2$ (normalized to 3). The CVC depicts a lack of current at $eV < 2\Delta$ (dashed linear dependence on Fig. 13 is shown for comparison), and the absence of Josephson supercurrent at V=0 can be explained by the influence of uncompensated magnetic moments of the iron atoms. Tunnel peaks with a non-classic split are visible on the CVC derivative; this shape can be a consequence of the ~25% anisotropy in the k-space. It is easy to identify extremes in the distribution of the order parameter $\Delta \approx 3.4$–$4.3$ meV which correspond to the characteristic ratio $2\Delta/k_BT_c \approx 4.4$–$5.5$, similar to $2\Delta_L/k_BT_c$ for a large gap, estimated by us in 122-selenide $(K_{0.7}Na_{0.3})_xFe_{2-y}Se_2$ with $T_c \approx 33$ K.

Compare the fine structure of the tunnel peaks and the Andreev minima observed in the spectrum of the SnS contacts, obtained for the same sample. The inset on Fig. 13 shows fragments of dI(V)/dV-characteristics of stack contacts: the above-mentioned SIS array (m=3 contacts in a stack) and SnS array (m=2) containing the main gap singularities located at eV = $2\Delta$. We can clearly see that after normalization we get the reproduction of not only the positions of these features and their widths, but also a complex (triplet) fine structure: obviously, the angular distribution of the gap is quite difficult and cannot be described by any of the options for $\Delta(\theta)$ taken during the calculations in Fig. 4. Nevertheless, the reproduced fine structure observable in the spectra of the stacks with differing numbers of contacts m and area, turns out to be an intrinsic (bulk) property of the material and is not caused by the dimensional effects or some random factors inherent to the surface.

Let us now consider the experimental data on superconducting oxypnictides (system 1111). Fig. 14 shows the dynamic conductance spectrum (black color) of a single contact in a $GdO_{1-x}F_xFeAs$ sample with a critical temperature $T_c \approx 50K$ (Ref. 141) measured at T=4.2 K. A series of singularities is clearly visible; those that are more intense, are located at $V_{nL=1} \approx \pm 22$ mV, and those similar in shape but with a smaller amplitude at $V_{nL=2} \approx \pm 11$mV (marked in Fig. 14 as $n_L$=1 and $n_L$=2, respectively); according to Eq. (1), these minima determine the value of the large gap $\Delta_L \approx 11$ meV. The next feature located at lower biases, $V_{nS=1} \approx \pm 5$mV, is not a third Andreev subharmonic of the large gap (expected at $V_{nL=3} \approx \pm 7.3$ mV), and can be interpreted as $n_S$=1 feature from the small gap $\Delta_S \approx 2.5$ meV. Taking into account the possibility of forming arrays, it should be noted that the spectrum of a single contact is interpreted unambiguously. Taking the sample $T_c \approx 50K$ for the sake of assessment, we can calculate the characteristic ratio for the large gap $2\Delta_L/k_BT_c \approx 5.1$. If we assume that this dI(V)/dV-spectrum corresponds to a stack contact with m≥2, then the characteristic ratio would be $2\Delta_L/k_BT_c \approx 5.1/m \leq 2.55$, which for obvious reasons is not possible for a "leading" gap, since it turns out to be less than the BCS limit, 3.52. This means that we can use the resulting characteristic ratio for the correct normalization of the array spectra (and the determination of the appropriate m) in GdO(F)FeAs samples with close $T_c$.

The inset of Fig. 14 also shows the dI(V)/dV-spectrum of another Andreev junction, obtained in an optimally doped GdO(F)FeAs. A pronounced minimum in the dynamic conductance, located at V ≈ ±44 mV, is similar in shape to the first Andreev subharmonic from

$\Delta_L$ in the single-junction spectrum considered above. At V ≈ ±22mV one can observe the second singularity, and its position coincides with the $n_L$=1 minimum in the spectrum of a single SnS contact. Since a twofold change of the large gap magnitude is not possible at similar $T_c$, we will assume that the scaling of the spectrum bias to an integer number occurred due to the IMARE: the array was formed on a step of the cryogenic cleft, implemented as an SnSnS structure. Indeed, after normalizing the bias axis of the given spectra by factor two (top dI(V)/dV-spectrum in Fig. 14) we achieved a coincidence in the positions of both the first ($n_L$=1) and the second ($n_L$=2) Andreev minima from the large gap, as well as the $2\Delta_S$-singularities ($n_S$=1). The normalized CVC of the stack contact has a pronounced foot at low biases, which confirms the excess current transport in the contract due to Andreev reflections.

We note that the values of both gaps, obtained by the Andreev and intrinsic Andreev spectroscopy using different samples (with similar $T_c$), are reproducible, and do not depend on the size or the resistance of the contacts.

Fig. 15 shows the normalized CVC and its derivative dI(V)/dV (at T=4.2K) for a stack (with a number of contacts m=3) in a samarium-based oxypnictide $Sm_{0.7}Th_{0.3}OFeAs$ [142] optimally doped with thorium, with $T_c$ ≈ 52 K. In order to observe multiple Andreev reflections the diameter of SnS contacts, as estimated in Refs. [74,75,77] must be about 2a = 20–60 nm, which is orders of magnitude less than the size of the crystallite in these samples [142] and the typical terrace width (see Fig. 2).

The data obtained for Sm-1111 are also typical for ballistic SnS contacts due to the pronounced foot at low biases (see Fig. 15). The dynamic conductance spectrum has two clearly defined, similar in shape and sufficiently sharp features that resemble minima at $V_{nL=1}$ ≈ ±23.2 mV and $V_{nL=2}$ ≈ ±12.4 mV, as well as features at $V_{nL=3}$ ≈ ±8.4 mV. The positions of these three singularities correspond to Eq. (1) and form a linear dependence on the inverse subharmonic number 1/n that goes through the origin (as shown in the bottom inset of Fig. 15 by a gray solid circles). The slope of the line $V_{nL}$(1/n) determines the value of the large gap $\Delta_L$ ≈ 12.4 meV. The Andreev minimum with $n_L$=4 has a negligible amplitude since backgrounded with the sharp exponential rise of the dynamic conductance curve (which corresponds to the foot area on the CVC).

The subharmonic structure of the small gap $\Delta_S$ begins with intense dips when $V_{nS=1}$ ≈ ±4.9 mV, followed by the minima located at $V_{nS=2}$ ≈ ±2.7 mV (marked in Fig. 15 as black arrows). The doublet nature of the features corresponding to $2\Delta_S$ can be a result of the anisotropy of the small gap in the k-space. Since in oxypnictides, according to our data [75], $\Delta^L/\Delta_S$ ≈ 4.5, the SGS of the small gap is usually located in the foot region, formed by the excess transport through the band with the gap $\Delta_L$, which makes it difficult to observe the $\Delta_S$-minima. For clarity, the top inset in Fig. 15 shows a fragment of the dI(V)/dV-spectrum of the given contact at low bias volt-ages with a suppressed background containing the SGS of the small gap, which made it possible to resolve the third subharmonic ($n_S=3$) from $\Delta_S$. The dependence $V_{nS}(1/n)$ (bottom inset on Fig. 15, open circles) can be used to determine the small gap as 2.7 meV. It is obvious that the features of the dI(V)/dV-spectrum on Fig. 15 uniquely define two independent SGS: their positions group into two linear dependences $V_{nL,S}(1/n)$, here the minima $n_S=1$ are much more pronounced and do not match the expected position of the fourth subharmonic $n_L=4$.

Two independent SGS are reproducibly observed in the dynamic conductance spectra of SnS contacts in superconductors from the 1111 family. Fig. 16 shows normalized CVC and their derivatives for two stacks (m=6), obtained using a similar Sm-1111 sample. Regardless of the different area and therefore the resistance of these contacts, the position of the Andreev minima for both the large ($n_L$ labels) and small gap (vertical arrows and $2\Delta_S$ label) remains unchanged. Moreover, the general shape of the SnS spectra for these contacts, obtained at distinct points of the cryogenic cleavage, is extremely similar. Therefore, here and in the spectra of other SnS contacts, the observed features cannot be caused by the influence of the dimensional or surface effects. The subharmonic structure from the small gap is rather blurred in Fig. 16: most likely, this is associated with the fact that the mean free path of the quasiparticles from the band in which $\Delta_S$ is developed, is much less than that of the carriers from the band with $\Delta_L$. Still, the $2\Delta_S$-features are clearly visible, and their position is also well reproduced and does not depend on m. The latter serves as a confirmation of the fact, these features are realized due to the Andreev reflection of the particles from bands with a bulk small gap.

The obtained values of $\Delta_L$ and $\Delta_S$ in Sm-1111 (see Figs. 15 and 16) are analogous to the amplitudes of the order parameters in Gd-1111 samples with a similar $T_c$ (see Fig. 14). This coincidence is not surprising, considering the similar structure of these oxypnictides (the difference lies only in the composition of the spacer layers, but the structure of the superconducting Fe-As blocks is unchanged), the value of the crystal lattice parameters [141,142], and the quasiparticle density of states at the Fermi level [143,144].

The pronounced SGS from a large gap, containing up to five subharmonics, was observed by us in the highest quality contacts based on LaO$_{1-x}$F$_x$FeAs with $T_c \approx 21$ K [145]. Normalized dI(V)/dV-characteristics at T=4.2K for a stack structure SnSnS (m=2) are shown in Fig. 17. Arrows and labels $n_L$ denote the SGS minima corresponding to the large gap $\Delta_L \approx 4.7$ meV; the dashed line and the labels nS indicate the SGS from the small gap $\Delta_S \approx 0.9$ meV. The inset shows the dependence of the biases $V_{nL,S}$ on 1/n for a large (solid circles) and small gap (open circles). Regardless of the fact that the first minimum of $\Delta_L$ is slightly shifted toward zero bias relative to the position of $2\Delta_L/e$, the combining of the features at $V_n \approx \pm 8$, $\pm 4.6$ mV and $\pm 3$ mV to a single SGS (corresponding to $\Delta_L \approx 4.7$ meV) is beyond doubt, considering the similarity between their complex asymmetric shapes and fine structure (which, most likely, is caused by ~20% anisotropy of $\Delta_L$ in the k-space). The characteristic ratio of BCS for a large gap $2\Delta_L/k_BT_c \approx 5.2$ is close to the value determined earlier for other oxypnictides based on samarium and gadolinium. The minima at $V_n \approx \pm1.7$ mV and $\pm0.9$ mV obviously do not fit in the direct proportionality of $V_{nL}(1/n)$ and therefore, constitute a second small gap SGS; though, as we can see at first glance (Fig. 17) due to the powerful $\Delta_L$ foot their amplitude is not very significant.

The experimental data examples considered above have shown that, despite the break-junction geometry is not known, the amplitudes of the superconducting gaps and the number of contacts in the stack (if the latter is implemented on a step of a cryogenic cleavage) can be reliably established using data statistics and comparing the reproducible features of the CVC and dI(V)/dV-spectra.

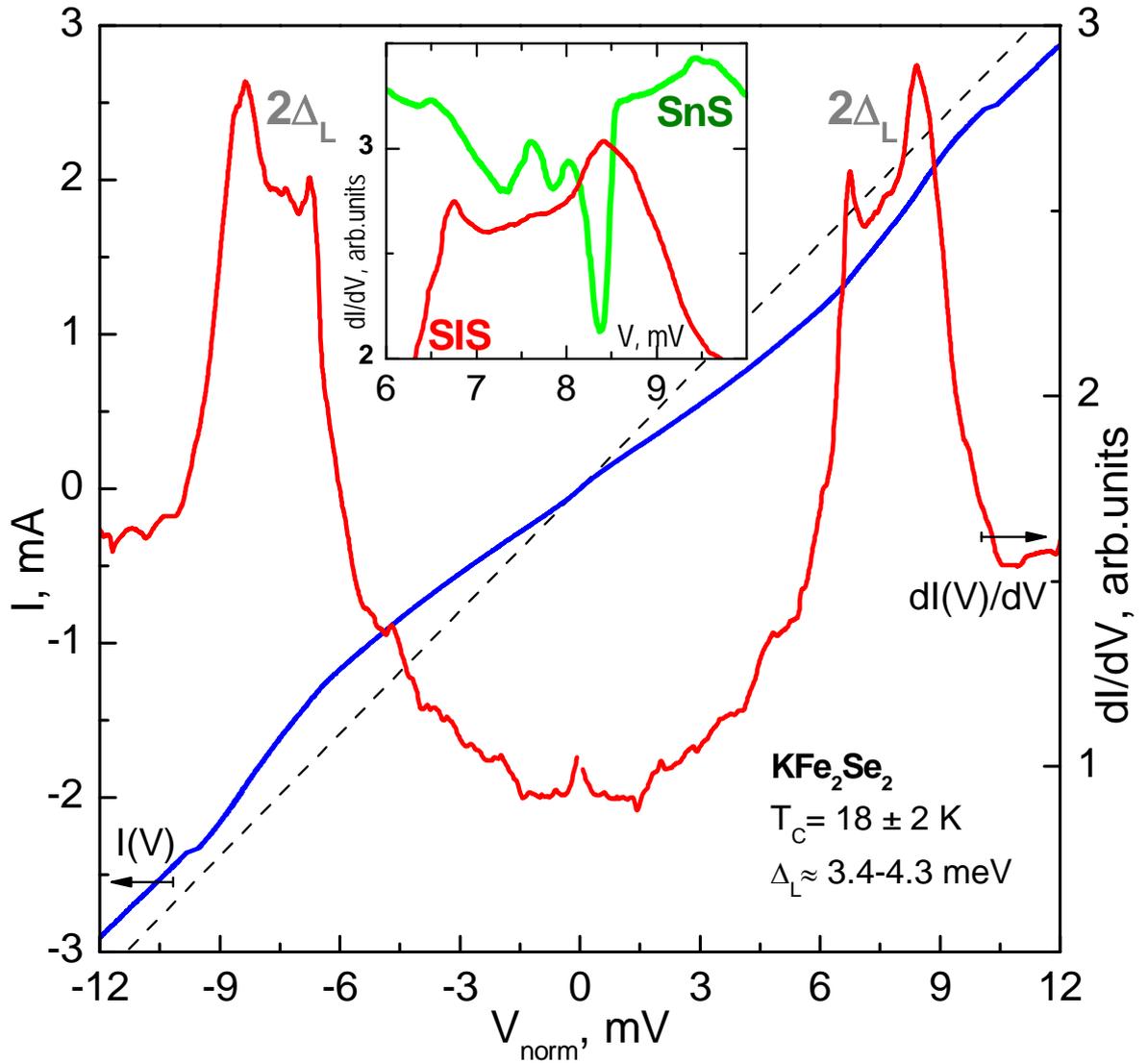

*Fig. 13. CVC normalized to that of a single contact (left vertical axis) and dI(V)/dV-spectrum (right axis) of a SIS array (m=3 contacts in a stack), obtained for a KFe$_2$Se$_2$ single crystal with $T_c \approx 18$ K. T=4.2 K. The positions of the doublet maxima determine the value of the gap $\Delta \approx 3.4$–4.3 meV (the range of values reflects the gap anisotropy ~25%). A linear dependence (dashed line) demonstrates the lack of current on the CVC and is shown for comparison. The inset shows a fragment of this spectrum (2Δ-maximum), as well as that of the SnS array (m=2 contacts in the stack), obtained for the same sample by sequential mechanical readjustment (the main Andreev minimum n=1). The bias voltage of both fragments is normalized to m=3 and 2, respectively; the position of the 2Δ-features (including a fine structure caused by anisotropy) coincides.*

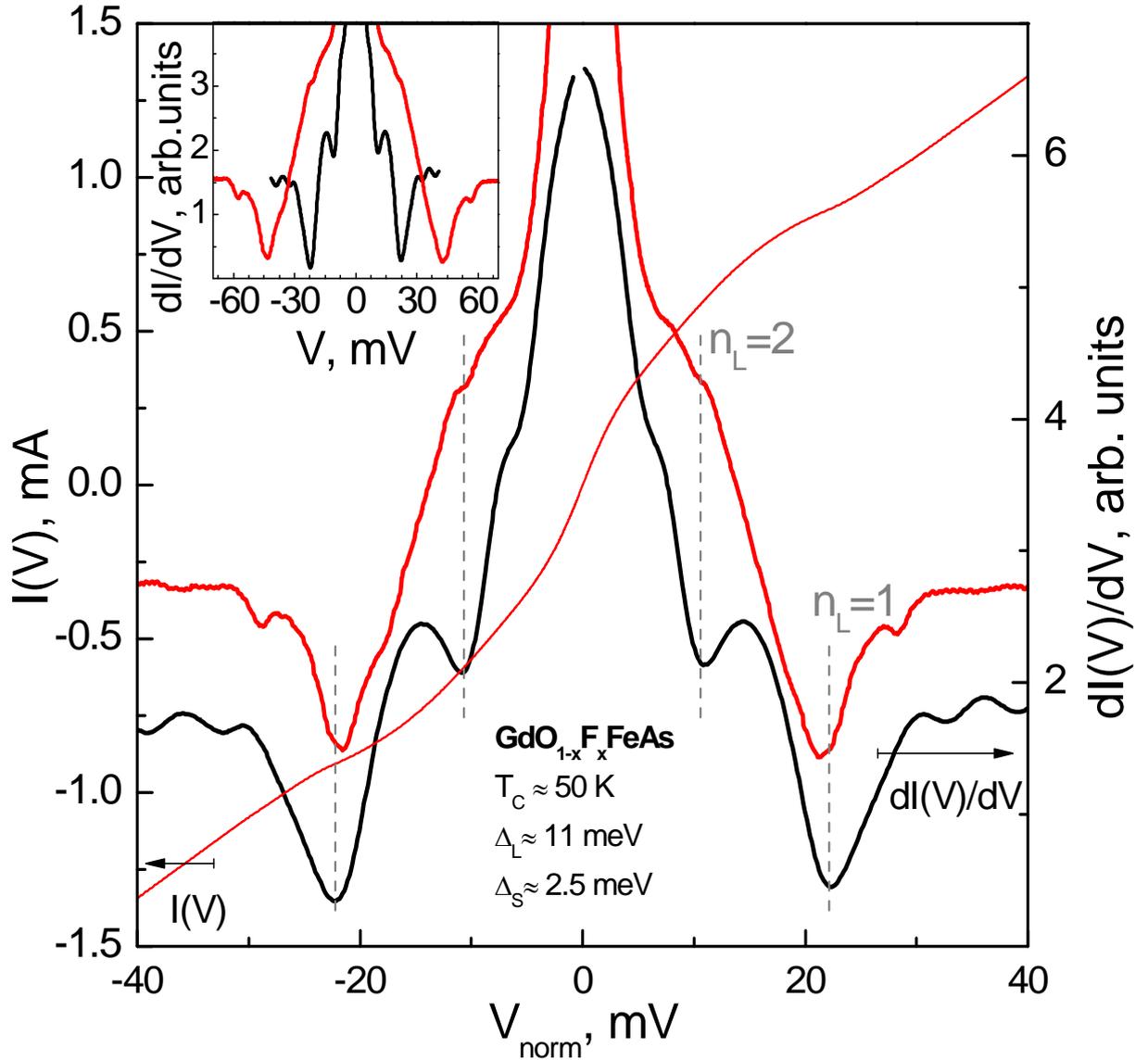

*Fig. 14. The dynamic conductance of a single SnS contact (black curve) as well as normalized (red curves) CVCs (left vertical axis) and dI(V)/dV-spectrum (right vertical axis) of a two-contact stack, obtained in optimally doped samples $GdO_{1-x}F_xFeAs$ with $T_c \approx 50$ K. T=4.2 K. The position of SGS for the large gap $\Delta_L \approx 11$ meV is shown by dashed lines and labels $n_L$=1,2. The inset shows the raw spectra before normalization.*

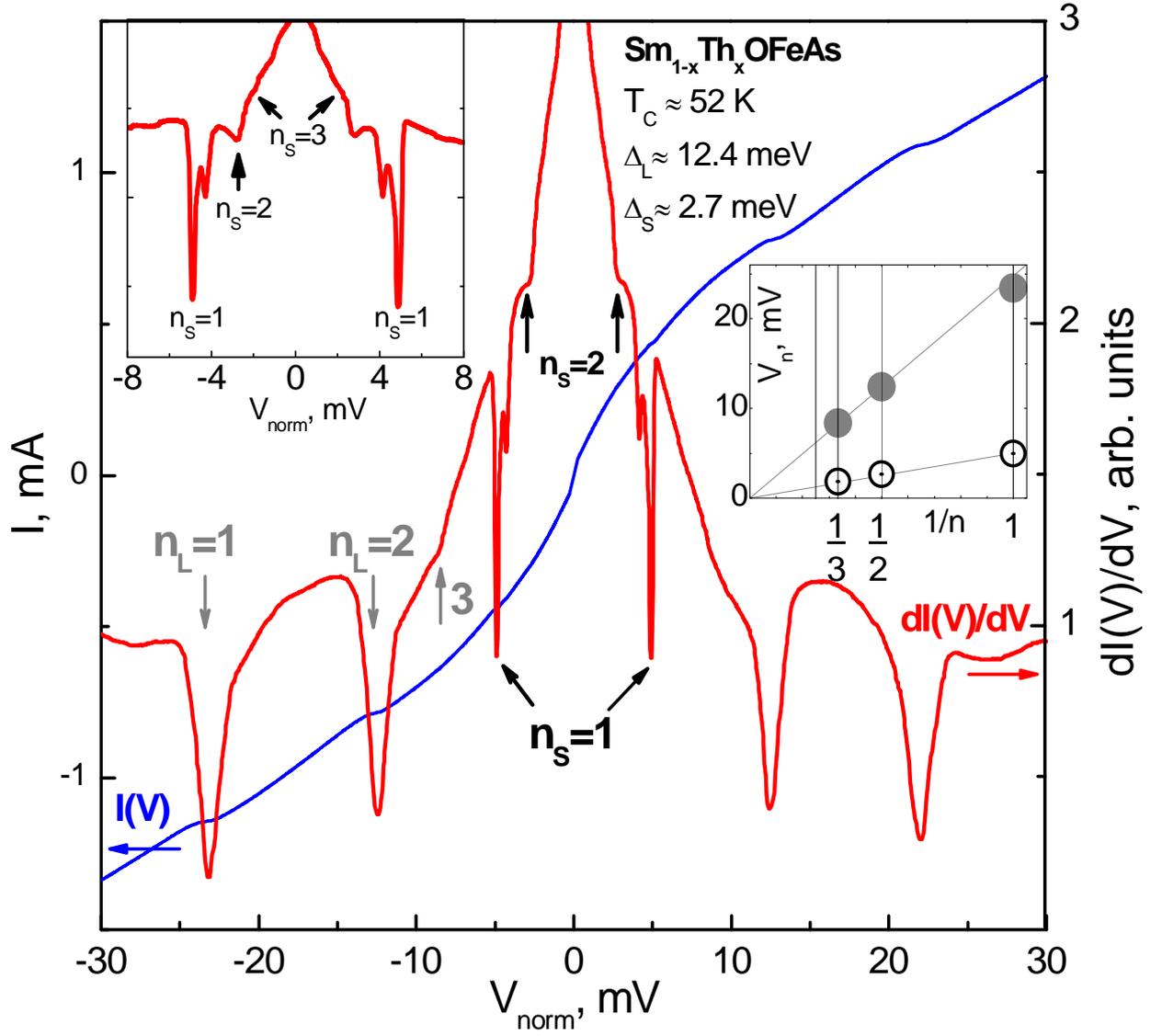

*Fig. 15.* CVC normalized to that of a single contact (left vertical axis) and the dynamic conductance (right axis) of an SnS stack (m=3 contacts) in a Sm$_{1-x}$Th$_x$OFeAs polycrystalline sample with $T_c \approx 52$ K. T=4.2 K. SGS of the large gap $\Delta_L \approx 12.4$ meV is shown in gray arrows and labels $n_L$=1,2,3; SGS from the small gap $\Delta_S \approx 2.7$ meV is noted by black arrows and labels $n_S$=1,2,3. The top inset shows a fragment of dI(V)/dV (with a suppressed background for clarity) containing SGS from a small gap. The bottom inset demonstrates the dependence of the Andreev minima positions Vn on 1/n for $\Delta_L$ (●) and $\Delta_S$ (○).

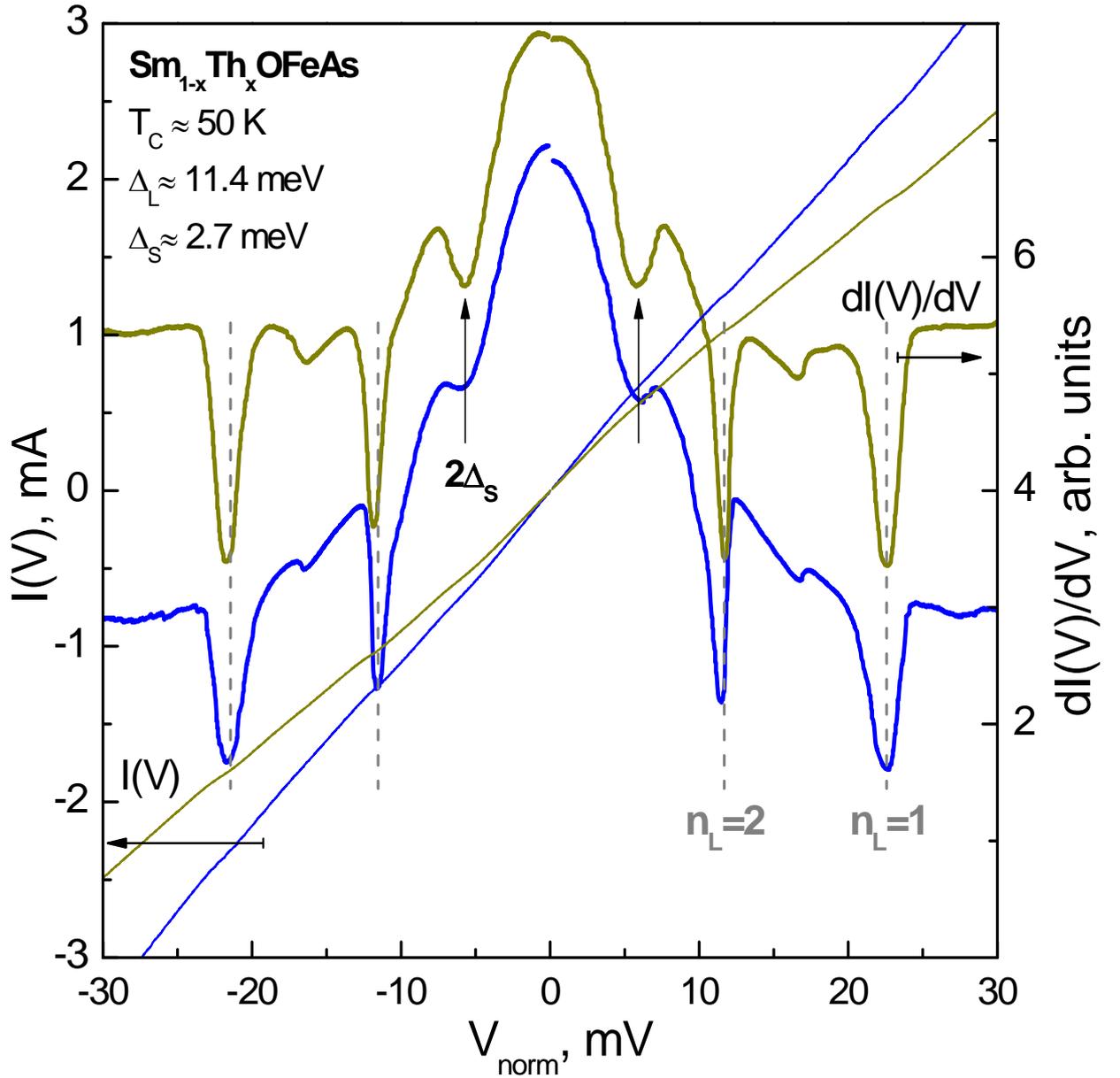

*Fig. 16. CVC normalized to that of a single contact (left vertical axis) and dynamic conductance spectra (right axis) of two Andreev arrays (m=6 SnS contacts per stack), created in an optimally doped $Sm_{1-x}Th_xOFeAs$ sample with $T_c \approx 50K$ by gentle mechanical readjustment. T=4.2 K. SGS from the large gap $\Delta_L \approx 11.4$ meV is denoted by dashed lines and labels $n_L=1,2$, the main Andreev minima nS¼1 from the small gap $\Delta_S \approx 2.7$ meV are shown by arrows and the label $2\Delta_S$.*

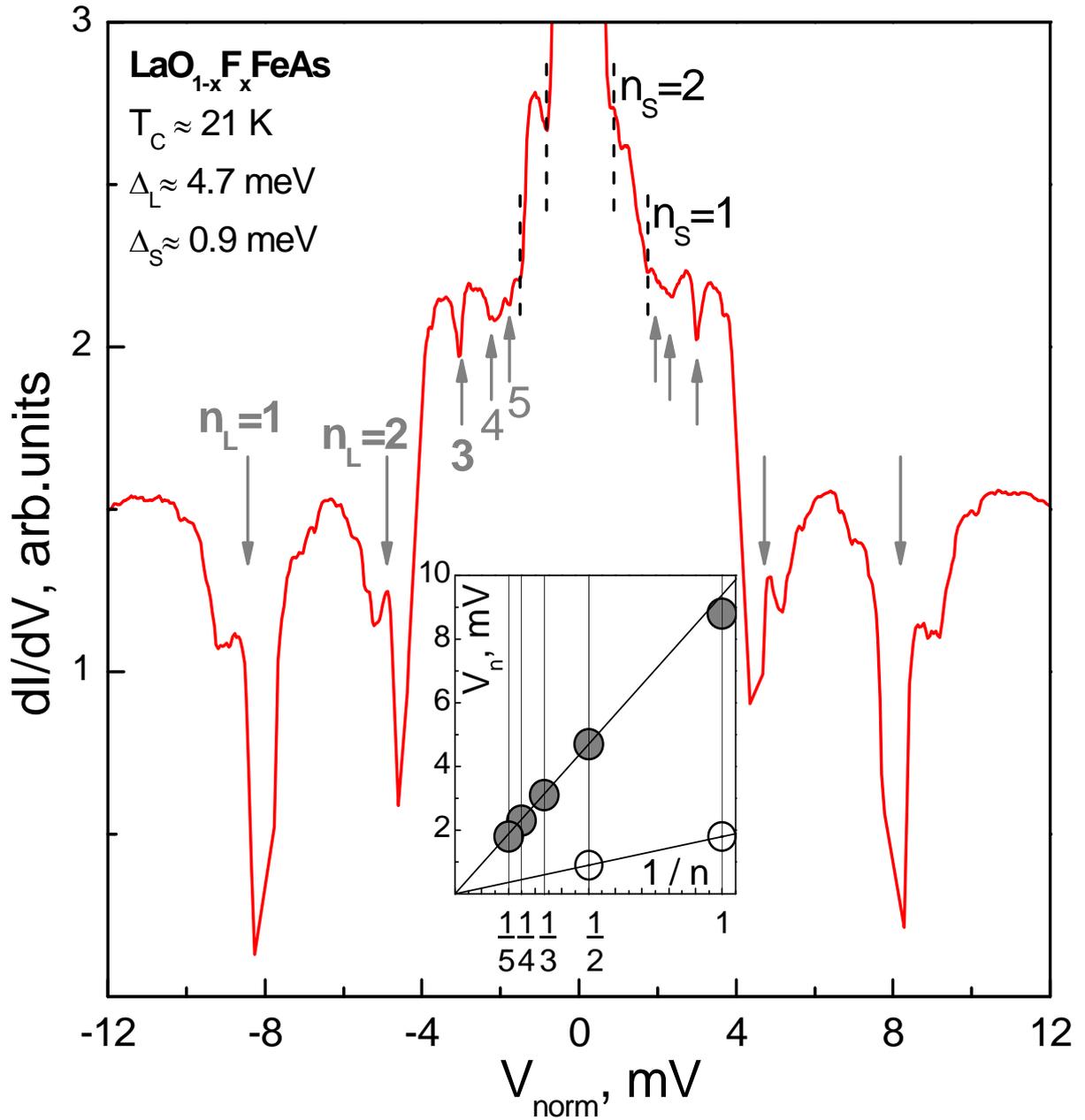

*Fig. 17.* A normalized dI(V)/dV-spectrum of an Andreev stack (m=2 SnS contacts), obtained for LaO$_{1-x}$F$_x$FeAs with $T_c \approx 21$ K. T=4.2 K. SGS of the large gap $\Delta_L \approx 4.7$ meV is shown by gray arrows and labels $n_L$=1,2... SGS from the small gap $\Delta_S \approx 0.9$ meV is marked by a black dotted line and the labels $n_S$=1,2. The inset shows the dependence of the Andreev minima positions $V_n$ on $1/n$ for $\Delta_L$ (●) and $\Delta_S$ (○).

## 5. Conclusion

To summarize the above, the "break-junction" technique is a powerful and often indispensable tool in the fundamental study of superconducting properties. It successfully works with the single crystals of any superconductor (both isotropic and layered) and also in polycrystalline samples of layered compounds. At the same time, the average size of the crystal grains in a polycrystalline sample must be several times greater than the break-junction diameter, i.e., it must be greater than 100 nm. The general requirement for the samples is the presence of a single superconducting phase and a size of no less than 1.5·3 mm$^2$ in the ab-plane. The inhomogeneity of the samples is controlled by measuring the CVC and dI(V)/dV-spectra in the range of temperatures up to $T_c$, and determining the local critical temperature of the contact area according to the linearization of the dynamic conductance.

The "break-junction" technique works reliably with whisker single crystals. It is possible to achieve junctions in which the current flows in ab crystallographic plane using film samples [27]. Nevertheless, obtaining junctions with j⊥c on bulk layered samples is impossible.

Finally, we will concisely present the advantages of the "break-junction", in the study of superconducting properties:

1) the use of clean cryogenic surfaces in the bulk of the sample, which ensures good heat sink from the tunnel junction point;
(2) the elimination of studied area overheating and remoteness of the contact area from the current leads of the sample;
(3) connecting the sample using a true four-point scheme;
(4) local measurement of the bulk superconducting parameters (using arrays);
(5) applicability for both single crystals and polycrystalline samples of layered compounds;
(6) applicability to materials in which the superconducting phase naturally occupies just a few percent of volume (for example, for the class of iron selenides $AFe_2Se_2$, where A = Na, K, Rb);
(7) the possibility of obtaining dozens of single and stack contacts on the cleavages of one sample by precise mechanical readjustment;
(8) the absence of chemical and mechanical influence on the tunnel junction region during the investigation;
(9) no loss of dopant oxygen from cryogenic clefts in the cuprate HTSC;
(10) the direction of the current as j∥c for layered single crystals;
(11) the high quality of resulting tunnel junctions and reproducible results;
(12) the ability to resolve the fine structure of dI(V)/dV-spectra, which allows to study the gap anisotropy and electron-phonon interaction;

(13) the possibility of implementing up to four methods of spectroscopy (Andreev and intrinsic Andreev, tunnel and intrinsic tunnel) on the same sample;

(14) the ability to resonantly excite bosons (optical phonons) with energies less than $4\Delta$ in the case of obtaining a SIS contact with alternating Josephson current;

(15) the ability to directly determine the amplitude of the order parameter in multi-gap superconductors based on the positions of the Andreev minima as $\langle\Delta_i\rangle = \langle eV_{n,i}\cdot n/2\rangle$ at any temperature up to $T_c$, in the case of an Andreev SnS contact (high transparency), which allows us to accurately evaluates the set of electron-boson coupling constants $\lambda_{ij}$.

In recent years the "break-junction" technique has shed light on the controversial and poorly understood aspects of the superconductivity mechanism in layered compounds, which is a key problem for condensed matter physics. To summarize HTSC studies we will conclude that the characteristic BCS ratio (which indirectly manifests the electron-boson interaction strength) does not depend notably on the degree of doping, and therefore, $T_c$. In particular, for cuprates this means that the carrier coupling mechanisms does not change during transitions from underdoped to overdoped regions. The study of two-gap superconductors that have been discovered thus far shows that for all of them the interband coupling strength is inferior to the intraband, which indicates that the latter plays the decisive role. Like any other method of research, the "break-junction" has its advantages and disadvantages, however its unique strength is the precise determination of the characteristic ratio $2\Delta/k_BT_c$. This fact should serve as a starting point for the construction of theoretical models describing the HTSC phenomenon.

The authors wish to thank V. M. Pudalov for the useful discussions; A. Krapf, W. Kraak, D. Wehler, H. Piel, L. I. Leonyuk, T. E. Os'kina, S. I. Krasnosvobodtsev, B. M. Bulychev, K. P. Burdina, O. V. Kravchenko, L. G. Sevastyanova, L. F. Kulikova, E. P. Khlybov, N. D. Zhigadlo, B. B€uchner, I. V. Morozov, M. V. Roslova, A. N. Vasiliev, and K. S. Perfakov for providing materials and characterizing the samples. The study is supported by the Council of the President of the Russian Federation for support of young scientists No. MK-5699.2016.2.

**Literature**

[1] J. Bardeen, L. N. Cooper, and J. R. Schrieffer, Phys. Rev. 108, 1175 (1957).

[2] B. D. Josephson, Phys. Lett. 1, 251 (1962).

[3] A. F. Andreev, JETP 46, 1823 (1964).

[4] R. C. Dynes, J. P. Garno, G. B. Hertel, and T. P. Orlando, Phys. Rev. Lett. 53, 2437 (1984).

[5] G. E. Blonder, M. Tinkham, and T. M. Klapwijk, Phys. Rev. B 25, 4515 (1982).

[6] M. Octavio, M. Tinkham, G. E. Blonder, and T. M. Klapwijk, Phys. Rev. B 27, 6739 (1983).

[7] I. K. Yanson, V. V. Fisun, N. L. Bobrov, Yu. G. Naidyuk, W. N. Kang, E.-M. Choi, H.-J. Kim, and S.-I. Lee, Phys. Rev. B 67, 024517 (2003).

[8] Ya. G. Ponomarev, E. B. Tsokur, M. V. Sudakova, S. N. Tchesnokov, M. E. Shabalin, M. A. Lorenz, M. A. Hein, G. Müller, H. Piel, and B. A. Aminov, Solid State Commun. 111, 513 (1999).

[9] G. Binnig, H. Rohrer, Ch. Gerber, and E. Weibel, Appl. Phys. Lett. 40, 178 (1982).

[10] G. Binnig and H. Rohrer, Physica B 127, 37 (1984).

[11] Tunneling Spectroscopy, edited by P. K. Hansma (Plenum Press, New York, 1982).

[12] E. L. Wolf, J. Zasadzinski, J. W. Osmun, and G. B. Arnold, J. Low Temp. Phys. 40, 19 (1980).

[13] E. L. Wolf, Principles of Electron Tunneling Spectroscopy, 2nd ed. (Oxford Science Publications, 2011).

[14] J. Moreland, S. Alexander, M. Cox, R. Sonnenfeld, and P. K. Hansma, Appl. Phys. Lett. 43, 387 (1983).

[15] J. Moreland and P. K. Hansma, Rev. Sci. Instrum. 55, 399 (1984).

[16] J. C. Cuevas and E. Scheer, Molecular Electronics (World Scientific, Singapore, 2010).

[17] P. K. Hansma, IBM J. Res. Dev. 30, 370 (1986).

[18] J. Moreland and J. W. Ekin, Appl. Phys. Lett. 47, 175 (1985).

[19] I. K. Yanson, Sov. Phys. JETP 39, 506 (1974).

[20] D. Daghero and R. S. Gonnelli, Supercond. Sci. Technol. 23, 043001 (2010).

[21] F. Giubileo, D. Roditchev, W. Sacks, R. Lamy, D. X. Thanh, J. Klein, S. Miraglia, D. Fruchart, J. Marcus, and Ph. Monod, Phys. Rev. Lett. 87, 177008 (2001).

[22] D. Roditchev, F. Giubileo, F. Bobba, R. Lamy, E.-M. Choi, H.-J. Kim, W. N. Kang, S. Miragliad, J. Marcus, W. Sacks, J. Klein, A. M. Cucolo, S.-I. Lee, and D. Fruchart, Physica C 408–410, 768 (2004).

[23] C. J. Müller, J. M. van Ruitenbeek, and L. J. de Jongh, Physica C 191, 485 (1992).

[24] J. Moreland and J. W. Ekin, J. Appl. Phys. 58, 3888 (1985).

[25] Th. Becherer, J. Kowalewski, M. Schmitt, M. Huth, W. Assmus, and H. Adrian, Z. Phys. B Condens. Mater. 86, 23 (1992).

[26] J. S. Tsai, I. Takeuchi, J. Fujita, T. Yoshitake, S. Miura, S. Tanaka, T. Terashima, Y. Bando, K. Iijima, and K. Yamamoto, Physica C 153–155, 1385 (1988).

[27] R. J. P. Keijsers, O. I. Shklyarevskii, J. G. H. Hermsen, and H. van Kempen, Rev. Sci. Instrum. 67, 2863 (1996).

[28] B. A. Aminov, A. I. Akimov, I. B. Brandt, M. T. Ngyuen, M. V. Sudakov, Yu. A. Pirogov, and Ya. G. Ponomarev, Sov. J. Low Temp. Phys. 15, 689 (1989).

[29] Ya. G. Ponomarev, N. B. Brandt, C. S. Khi, S. V. Tchesnokov, E. B. Tsokur, A. V. Yarygin, K. T. Yusupov, B. A. Aminov, M. A. Hein, G. M€uller, H. Piel, D. Wehler, V. Z. Kresin, K. Rosner, K. Winzer, and Th. Wolf, Phys. Rev. B 52, 1352 (1995).

[30] Ya. G. Ponomarev, H. H. Van, S. A. Kuzmichev, S. V. Kulbachinskii, M. G. Mikheev, M. V. Sudakova, and S. N. Tchesnokov, JETP Lett. 96, 743 (2013).

[31] Ya. G. Ponomarev, B. A. Aminov, M. A. Hein, H. Heinrichs, V. Z. Kresin, G. M€uller, H. Piel, K. Rosner, S. V. Tchesnokov, E. B. Tsokur, D. Wehler, K. Winzer, A. V. Yarygin, and K. T. Yusupov, Physica C 243, 167 (1995).

[32] J. Moreland, J. W. Ekin, L. F. Goodrich, T. E. Capobianco, A. F. Clark, J. Kwo, M. Hong, and S. H. Liou, Phys. Rev. B 35, 8856 (1987).

[33] J. Moreland, A. F. Clark, L. F. Goodrich, H. C. Ku, and R. N. Shelton, Phys. Rev. B 35, 8711 (1987).

[34] T. Ekino, A. M. Gabovich, M. S. Li, M. Pekala, H. Szymczak, and A. I. Voitenko, Physica C 468, 1145 (2008).

[35] N. Miyakawa, J. F. Zasadzinski, S. Oonuki, M. Asano, D. Henmi, T. Kaneko, L. Ozyuzer, and K. E. Gray, Physica C 364–365, 475 (2001).

[36] Ya. G. Ponomarev, Phys. Usp. 45, 649 (2002).

[37] Ya. G. Ponomarev and E. G. Maksimov, JETP Lett. 76, 394 (2002).

38] S. I. Vedeneev, A. A. Tsvetkov, A. G. M. Jansen, and P. Wyder, Physica C 235–240, 1851 (1994).

[39] R. S. Gonnelli, G. A. Ummarino, and V. A. Stepanov, Physica C 282–287, 1473 (1997)

[40] D. K. Petrov, Ya. G. Ponomarev, Kh. T. Rahimov, K. Steupati, M. V. Sudakova, A. B. Tennakun, and A. D. Shevchenko, Sov. J. Low Temp. Phys. 17, 445 (1991).

[41] B. Aminov, L. Roschta, Ya. G. Ponomarev, and M. V. Sudakova, Sov. J. Low Temp. Phys. 17, 364 (1991).

[42] B. A. Aminov, M. A. Hein, M. A. Lorenz, G. M€uller, H. Piel, D. Wehler, V. Z. Kresin, Ya. G. Ponomarev, I. A. Borisova, C. S. Chi, E. B. Tsokur, L. Buschmann, L. Winkeler, G. Güntherodt, and K. Winzer, J. Low. Temp. Phys. 105, 1225 (1996).

[43] Ya. G. Ponomarev, C. S. Khi, K. K. Uk, M. V. Sudakova, S. N. Tchesnokov, M. A. Lorenz, M. A. Hein, G. M€uller, H. Piel, B. A. Aminov, A. Krapf, and W. Kraak, Physica C 315, 85 (1999).

[44] B. A. Aminov, D. Wehler, G. M€uller, H. Piel, M. A. Hein, H. Heinrichs, N. B. Brandt, C. S. Hu, Y. G. Ponomarev, E. B. Tsokur, S. N. Chesnokov, K. C. Yusupov, A. V. Yarygin, K. Winzer, K. Rosner, and T. Wolf, JETP Lett. 60, 424 (1994).

[45] G. Krabbes, R. M€uller, M. Ritschel, H. Vinzelberg, E. Wolf, B. A. Aminov, M. T. Nguen, M. V. Sudakova, and Ya. G. Ponomarev, Phys. Status Solidi 104, K61 (1987).

[46] J. R. Kirtley and F. Tafuri, Handbook of High-Temperature Superconductivity, edited by J. R. Schrieffer (2007), p. 19.

[47] J. R. Kirtley, Int. J. Mod. Phys. B 4, 201 (1990).

[48] Ya. G. Ponomarev, S. A. Kuzmichev, N. M. Kadomtseva, M. G. Mikheev, M. V. Sudakova, S. N. Chesnokov, E. G. Maksimov, S. I. Krasnosvobodtsev, L. G. Sevast'yanova, K. P. Burdina, and B. M. Bulychev, JETP Lett. 79, 484 (2004).

[49] Ya. G. Ponomarev, S. A. Kuzmichev, M. G. Mikheev, M. V. Sudakova, S. N. Tchesnokov, N. Z. Timergaleev, A. V. Yarigin, E. G. Maksimov, S. I. Krasnosvobodtsev, A. V. Varlashkin, M. A. Hein, G. M€uller, H. Piel, L. G. Sevastyanova, O. V. Kravchenko, K. P. Burdina, and B. M. Bulychev, Solid State Commun. 129, 85 (2004).

[50] Ya. G. Ponomarev, S. A. Kuzmichev, M. G. Mikheev, M. V. Sudakova, S. N. Tchesnokov, H. Van Hoai, B. M. Bulychev, E. G. Maksimov, and S. I. Krasnosvobodtsev, JETP Lett. 85, 46 (2007).

[51] S. A. Kuzmichev, T. E. Shanygina, S. N. Tchesnokov, and S. I. Krasnosvobodtsev, Solid State Commun. 152, 119 (2012).

[52] S. A. Kuzmichev, T. E. Kuzmicheva, and S. N. Tchesnokov, JETP Lett. 99, 295 (2014).

[53] S. A. Kuzmichev, T. E. Kuzmicheva, S. N. Tchesnokov, V. M. Pudalov, and A. N. Vasiliev, J. Supercond. Novel Magn. 29, 1111 (2016).

[54] T. Ekino, T. Takasaki, T. Muranaka, J. Akimitsu, and H. Fujii, Phys. Rev. B 67, 094504 (2003).

[55] T. Ekino, A. M. Gabovich, M. S. Li, T. Takasaki, A. I. Voitenko, J. Akimitsu, H. Fujii, T. Muranaka, M. Pekala, and H. Szymczak, Physica B 359–361, 460 (2005).

[56] R. A. Ribeiro, T. Ekino, T. Takasaki, T. Takabatake, and J. Akimitsu, Physica C 426–431, 450 (2005).

[57] R. S. Gonnelli, G. A. Ummarino, D. Daghero, A. Calzolari, and V. A. Stepanov, Int. J. Mod. Phys. B 16, 1553 (2002).

[58] R. S. Gonnelli, A. Calzolari, D. Daghero, G. A. Ummarino, V. A. Stepanov, G. Giunchi, S. Ceresara, and G. Ripamonti, Phys. Rev. Lett. 87, 097001 (2001).

[59] Z.-Z. Li, Y. Xuan, H.-J. Tao, P.-S. Luo, Z.-A. Ren, G.-C. Che, B.-R. Zhao, and Z.-X. Zhao, Physica C 370, 1 (2002).

[60] H. Schmidt, J. F. Zasadzinski, K. E. Gray, and D. G. Hinks, Physica C 385, 221 (2003).

[61] T. Takasaki, T. Ekino, A. M. Gabovich, A. Sugimoto, S. Yamanaka, and J. Akimitsu, Superconductivity, edited by V. R. Romanovskii (2012), Chap. 1.

[62] A. Sugimoto, T. Ekino, R. Ukita, K. Shohara, H. Okabe, J. Akimitsu, and A. M. Gabovich, Physica C 470, 1070 (2010).

[63] T. Ekino, A. Sugimoto, H. Okabe, K. Shohara, R. Ukita, J. Akimitsu, and A. M. Gabovich, Physica C 470, S358 (2010).

[64] T. E. Shanygina, Ya. G. Ponomarev, S. A. Kuzmichev, M. G. Mikheev, S. N. Tchesnokov, O. E. Omel'yanovskii, A. V. Sadakov, Yu. F. Eltsev, A. S. Dormidontov, V. M. Pudalov, A. S. Usol'tsev, and E. P. Khlybov, JETP Lett. 93, 94 (2011).

[65] V. M. Pudalov, O. E. Omelyanovskiy, E. P. Khlybov, A. V. Sadakov, Yu. F. Eltsev, K. V. Mitsen, O. M. Ivanenko, K. S. Pervakov, D. R. Gizatulin, A. S. Ysoltsev, A. S. Dormidontov, S. Yu. Gavrilkin, A. Yu. Tsvetkov, Ya. G. Ponomarev, S. A. Kuzmichev, M. G. Mikheev, S. N. Chesnokov, T. E. Shanygina, and S. M. Kazakov, Phys. Usp. 54, 648 (2011).

[66] Ya. G. Ponomarev, S. A. Kuzmichev, M. G. Mikheev, M. V. Sudakova, S. N. Tchesnokov, T. E. Shanygina, O. S. Volkova, A. N. Vasiliev, and Th. Wolf, J. Exp. Theor. Phys. 113, 459 (2011).

[67] S. A. Kuzmichev, T. E. Shanygina, I. V. Morozov, A. I. Boltalin, M. V. Roslova, S. Wurmehl, and B. B€uchner, JETP Lett. 95, 537 (2012).

[68] T. E. Shanygina, S. A. Kuzmichev, M. G. Mikheev, Y. G. Ponomarev, S. N. Tchesnokov, Y. F. Eltsev, V. M. Pudalov, A. V. Sadakov, A. S. Usol'tsev, E. P. Khlybov, and L. F. Kulikova, J. Supercond. Novel Magn. 26, 2661 (2013).

[69] Ya. G. Ponomarev, S. A. Kuzmichev, T. E. Kuzmicheva, M. G. Mikheev, M. V. Sudakova, S. N. Tchesnokov, O. S. Volkova, A. N. Vasiliev, V. M. Pudalov, A. V. Sadakov, A. S. Usol'tsev, Th. Wolf, and E. P. Khlybov, J. Supercond. Novel Magn. 26, 2867 (2013).

[70] D. Chareev, E. Osadchii, T. Kuzmicheva, Jiunn-Yuan Lin, S. Kuzmichev, O. Volkova, and A. Vasiliev, CrystEngComm 15, 1989 (2013).

[71] T. E. Kuzmicheva, S. A. Kuzmichev, M. G. Mikheev, Ya. G. Ponomarev, S. N. Tchesnokov, Yu. F. Eltsev, V. M. Pudalov, K. S. Pervakov, A. V. Sadakov, A. S. Usoltsev, E. P. Khlybov, and L. F. Kulikova, EPL 102, 67006 (2013).

[72] S. A. Kuzmichev, T. E. Kuzmicheva, A. I. Boltalin, and I. V. Morozov, JETP Lett. 98, 722 (2014).

[73] M. V. Roslova, S. Kuzmichev, T. Kuzmicheva, Y. Ovchenkov, Min Liu, I. Morozov, A. Boltalin, A. Shevelkov, D. Chareev, and A. Vasiliev, CrystEngComm 16, 6919 (2014).

[74] T. E. Kuzmicheva, S. A. Kuzmichev, and N. D. Zhigadlo, JETP Lett. 99, 136 (2014).

[75] T. E. Kuzmicheva, S. A. Kuzmichev, M. G. Mikheev, Ya. G. Ponomarev, S. N. Tchesnokov, V. M. Pudalov, E. P. Khlybov, and N. D. Zhigadlo, Phys. Usp., 57, 819 (2014).

[76] M. Abdel-Hafiez, P. J. Pereira, S. A. Kuzmichev, T. E. Kuzmicheva, V. M. Pudalov, L. Harnagea, A. A. Kordyuk, A. V. Silhanek, V. V. Moshchalkov, B. Shen, H.-H. Wen, A. N. Vasiliev, and X.-J. Chen, Phys. Rev. B 90, 054524 (2014).

[77] T. E. Kuzmicheva, S. A. Kuzmichev, S. N. Tchesnokov, and N. D. Zhigadlo, J. Supercond. Novel Magn. 29, 673 (2016).

[78] M. A. Reed, C. Zhou, C. J. Muller, T. P. Burgin, and J. M. Tour, Science 278, 252 (1997).

[79] D. Xiang, H. Jeong, T. Lee, and D. Mayer, Adv. Mater. 25, 4845 (2013).

[80] M. A. Lorenz, M. A. Hein, G. Müller, H. Piel, H. Schmidt, Y. G. Ponomarev, M. V. Sudakova, S. N. Tchesnokov, E. B. Tsokur, M. E. Shabalin, and B. A. Aminov, J. Low Temp. Phys. 117, 527 (1999).

[81] Ya. G. Ponomarev, K. K. Uk, and M. A. Lorenz, Inst. Phys. Conf. Ser. 167, 241 (2000).

[82] N. Agrait, J. G. Rodrigo, and S. Vieira, Phys. Rev. B 46, 5814 (1992).

[83] Y. Yin, M. Zech, T. L. Williams, and J. E. Hoffman, Physica C 469, 535 (2009).

[84] Yu. V. Sharvin, JETP 48, 984 (1965).

[85] G. B. Arnold, J. Low Temp. Phys. 68, 1 (1987).

[86] D. Averin and A. Bardas, Phys. Rev. Lett. 75, 1831 (1995).

[87] J. C. Cuevas, A. Martı́n-Rodero, and A. Levy Yeyat, Phys. Rev. B 54, 7366 (1996); A. Poenicke, J. C. Cuevas, and M. Fogelstrom, ibid. 65, 220510(R) (2002).

[88] R. Kümmel, U. Gunsenheimer, and R. Nicolsky, Phys. Rev. B 42, 3992 (1990).

[89] J. Nagamatsu, N. Nakagawa, T. Muranaka, Y. Zenitani, and J. Akimitsu, Nature 410, 63 (2001).

[90] I. K. Yanson and Yu. G. Naidyuk, Fiz. Nizk. Temp. 30 (2004) [Low Temp. Phys. 30, 261 (2004)].

[91] X. X. Xi, Rep. Progr. Phys. 71, 116501 (2008).

[92] M. H. Badr, M. Freamat, Y. Sushko, and K.-W. Ng, Phys. Rev. B 65, 184516 (2002).

[93] Zh.-Zh. Li, H.-J. Tao, Y. Xuan, Z.-A. Ren, G.-C. Che, and B.-R. Zhao, Phys. Rev. B 66, 064513 (2002).

[94] T. Takasaki, T. Ekino, R. A. Ribeiro, T. Muranaka, H. Fujii, and J. Akimitsu, Physica C 426–431, 300 (2005).

[95] T. Ekino, A. M. Gabovich, M. S. Li, T. Takasaki, A. I. Voitenko, J. Akimitsu, H. Fujii, T. Muranaka, M. Pekala, and H. Szymczak, Physica C 426–431, 230 (2005).

[96] V. A. Moskalenko, FMM 8, 503 (1959); Sov. Phys. Usp. 17, 450 (1974).

[97] H. Suhl, B. T. Matthias, and L. R. Walker, Phys. Rev. Lett. 3, 552 (1959).

[98] B. A. Aminov, M. A. Hein, G. M€uller, H. Piel, D. Wehler, Y. G. Ponomarev, K. Rosner, and K. Winzer, J. Supercond. 7, 361 (1994).

[99] V. Z. Kresin and S. A. Wolf, Phys. Rev. B 46, 6458 (1992).

[100] N. Klein, N. Tellmann, H. Schulz, K. Urban, S. A. Wolf, and V. Z. Kresin, Phys. Rev. Lett. 71, 3355 (1993).

[101] J. Bouvier and J. Bok, Physica C 249, 117 (1995).

[102] J. Bok and J. Bouvier, Physica C 274, 1 (1997).

[103] S. Onari, H. Kontani, and M. Sato, Phys. Rev. B 81, 060504(R) (2010).

[104] F. Ahn, I. Eremin, J. Knolle, V. B. Zabolotnyy, S. V. Borisenko, B. Büchner, and A. V. Chubukov, Phys. Rev. B 89, 144513 (2014).

[105] T. P. Devereaux and P. Fulde, Phys. Rev. B 47, 14638 (1993).

[106] R. Kleiner and P. M€uller, Phys. Rev. B 49, 1327 (1994); Physica C 293, 156 (1997).

[107] K. Schlenga, R. Kleiner, G. Hechtfischer, M. Mößle, S. Schmitt, Paul Müller, Ch. Helm, Ch. Preis, F. Forsthofer, J. Keller, H. L. Johnson, M. Veith, and E. Steinbeiß, Phys. Rev. B 57, 14518 (1998).

[108] A. A. Yurgens, Supercond. Sci. Technol. 13, R85 (2000).

[109] A. A. Abrikosov, Physica C 317–318, 154 (1999).

[110] B. A. Aminov, L. I. Leonyuk, T. E. Oskina, H. Piel, Y. G. Ponomarev, H. T. Rachimov, K. Sethupathi, M. V. Sudakova, and D. Wehler, Adv. Supercond. V, 1037 (1993).

[111] T. Yamashita, S.-J. Kim, Y. Latyshev, and K. Nakajima, Physica C 335, 219 (2000).

[112] V. M. Krasnov, N. Mros, A. Yurgens, and D. Winkler, Phys. Rev. B 59, 8463 (1999).

[113] Ya. G. Ponomarev, (unpublished).

[114] P. Nyhus, M. A. Karlow, S. L. Cooper, B. W. Veal, and A. P. Paulikas, Phys. Rev. B 50, 13898 (1994).

[115] X. H. Chen, K. Q. Ruan, G. G. Qian, S. Y. Li, L. Z. Cao, J. Zou, and C. Y. Xu, Phys. Rev. B 58, 5868 (1998).

[116] A. E. Pantoja, D. M. Pooke, H. J. Trodahl, and J. C. Irwin, Phys. Rev. B 58, 5219 (1998).

[117] U. Paltzer, F. W. de Wette, U. Schr€oder, and E. Rampf, Physica C 301, 55 (1998).

[118] E. G. Maksimov, Phys. Usp. 43, 965 (2000).

[119] N. Z. Timergaleev, PhD thesis, Physical Faculty, M.V. Lomonosov Moscow State University, (2002).

[120] J. E. Hirsch, Phys. Lett. A 282, 392 (2001).

[121] S. L. Bud'ko, G. Lapertot, C. Petrovic, C. E. Cunningham, N. Anderson, and P. C. Canfield, Phys. Rev. Lett. 86, 1877 (2001).

[122] D. G. Hinks, H. Claus, and J. D. Jorgensen, Nature 411, 457 (2001).

[123] J. Geerk, R. Schneider, G. Linker, A. G. Zaitsev, R. Heid, K.-P. Bohnen, and H. V. Löhneysen, Phys. Rev. Lett. 94, 227005 (2005).

[124] A. Floris, G. Profeta, N. N. Lathiotakis, M. L€uders, M. A. L. Marques, C. Franchini, E. K. U. Gross, A. Continenza, and S. Massidda, Phys. Rev. Lett. 94, 037004 (2005).

[125] A. Y. Liu, I. I. Mazin, and J. Kortus, Phys. Rev. Lett. 87, 087005 (2001).

[126] H. J. Choi, D. Roundy, H. Sun, M. L. Cohen, and S. G. Louie, Nature 418, 758 (2002); Phys. Rev. B 66, 020513 (2002).

[127] G. A. Ummarino, R. S. Gonnelli, S. Massidda, A. Bianconi, Physica C 407, 121 (2004).

[128] J. Kortus, O.V. Dolgov, R.K. Kremer, A. A. Golubov, Phys. Rev. Lett. 94, 027002 (2005).

[129] Yu. F. Eltsev, S. Lee, K. Nakao, N. Chikumoto, S. Tajima, N. Koshizuka, and M. Murakami, Phys. Rev. B 65, 140501(R) (2002).

[130] J. Karpinski, N. D. Zhigadlo, G. Schuck, S. M. Kazakov, B. Batlogg, K. Rogacki, R. Puzniak, J. Jun, E. Müller, P. Wägli, R. Gonnelli, D. Daghero, G. A. Ummarino, and V. A. Stepanov, Phys. Rev. B 71, 174506 (2005).

[131] L. G. Sevast'yanova, P. E. Kaz, O. V. Kravchenko, S. A. Kuz'michev, Ya. G. Ponomarev, K. P. Burdina, and B. M. Bulychev, Russ. Chem. Bull. 52, 1674 (2003).

[132] A. Brinkman, A. A. Golubov, H. Rogalla, O. V. Dolgov, J. Kortus, Y. Kong, O. Jepsen, and O. K. Andersen, Phys. Rev. B 65, 180517(R) (2002).

[133] S. Graser and T. Dahm, Phys. Rev. B 75, 014507 (2007).

[134] Y. Kamihara, T. Watanabe, M. Hirano, H. Hososno, J. Am. Chem. Soc. 130, 3296 (2008).

[135] P. J. Hirschfeld, M. M. Korshunov, and I. I. Mazin, Rep. Prog. Phys. 74, 124508 (2011).

[136] M. M. Korshunov, Phys. Usp. 57, 813 (2014).

[137] C. Wang, Z.-C. Wang, Y.-X. Mei, Y.-K. Li, L. Li, Z.-T. Tang, Y. Liu, P. Zhang, H.-F. Zhai, Z.-A. Xu, and G.-H. Cao, J. Am. Chem. Soc. 138, 2170 (2016).

[138] D. J. Singh, J. Alloys Compd. 687, 786 (2016).

[139] T. Saito, S. Onari, and H. Kontani, Phys. Rev. B 88, 045115 (2013).

[140] E. Dagotto, Rev. Mod. Phys. 85, 849 (2013).

[141] E. P. Khlybov, O. E. Omelyanovsky, A. Zaleski, A. V. Sadakov, D. R. Gizatulin, L. F. Kulikova, I. E. Kostyleva, and V. M. Pudalov, JETP Lett. 90, 387 (2009).

[142] N. D. Zhigadlo, S. Katrych, S. Weyeneth, R. Puzniak, P. J. W. Moll, Z. Bukowski, J. Karpinski, H. Keller, and B. Batlogg, Phys. Rev. B 82, 064517 (2010); 86, 214509 (2012).

[143] I. A. Nekrasov, Z. V. Pchelkina, and M. V. Sadovskii, JETP Lett. 87, 560 (2008).

[144] E. Z. Kuchinskii, I. A. Nekrasov, and M. V. Sadovskii, JETP Lett. 91, 518 (2010).

[145] A. Kondrat, J. E. Hamann-Borrero, N. Leps, M. Kosmala, O. Schumann, A. Köhler, J. Werner, G. Behr, M. Braden, R. Klingeler, B. Büchner, and C. Hess1a, Eur. Phys. J. B 70, 461 (2009).